%% file: main.tex
\documentclass[11pt]{article}

\usepackage[table]{xcolor} 
\definecolor{gray}{RGB}{230,230,230} 
\definecolor{cblue}{RGB}{173,213,232} 
\definecolor{cred}{RGB}{239, 170, 170}
\definecolor{cgreen}{RGB}{165, 199, 165}
\usepackage{tikz}
\usetikzlibrary{
    arrows.meta,         
    decorations.pathreplacing,  
    positioning          
}
\usetikzlibrary{arrows.meta, positioning}
\usepackage[T1]{fontenc}
\usepackage[utf8]{inputenc}
\usepackage[french]{babel}
\usepackage{blindtext}
\usepackage{amsthm}
\usepackage{amsmath}
\usepackage{booktabs}
\usepackage{amssymb}
\usepackage{graphicx}

\usepackage{geometry}
\usepackage{hyperref}
\usepackage{bbold}
\usepackage{mathtools}
\usepackage{multicol}
\usepackage{ragged2e}
\usepackage{tikz}
\usepackage{setspace}
\usepackage{algorithm}
 \usepackage{stmaryrd} 
\usepackage{algpseudocode} 
\usepackage{enumitem}  

\usepackage{longtable}
\usepackage{array}
\usepackage{comment}

\renewcommand{\arraystretch}{1} 
\usepackage{tabularx}
\usepackage{ulem}

 \usepackage{url}
 \usepackage{array}

 \usepackage{authblk} 

\usepackage[justification=centering]{caption}
\usetikzlibrary{arrows, shapes}

\usepackage{caption}
\usepackage{subcaption}

\newtheoremstyle{brakets}
{}
{}
{\itshape}
{}
{\bfseries}
{.}
{ }
{\thmheadbrackets{#1}{#2}{#3}}

\newtheorem{definition}{Definition}

\newtheorem{rmk}{Remark}

\newcommand{\R}{\mathbb{R}}
\newcommand{\N}{\mathbb{N}}
\usepackage{authblk}

\theoremstyle{definition}

 \title{\textbf{Clustering methods for categorical time series and sequences : a scoping review}}

\author[1,*]{Ottavio Khalifa}
\author[1]{Alan Balendran}
\author[1,2]{Viet-Thi Tran}
\author[1]{François Petit}

\affil[1]{Université Paris Cité, Université Sorbonne Paris Nord, INSERM, INRAE, Centre for Research in Epidemiology and StatisticS (CRESS), Paris, France.}

\affil[2]{Centre D'Epidemiologie Clinique, AP-HP, Hôpital Hôtel Dieu, F-75004, Paris, France}

\affil[*]{Corresponding author: ottavio.khalifa@inserm.fr}
\date{}
\begin{document}


\maketitle

\textbf{Objective:}  
To provide a comprehensive overview and typology of clustering methods for categorical time series (CTS), a data structure commonly found in epidemiology, sociology, biology, and marketing, and to support method selection in regards to data characteristics. \\

\textbf{Materials and Methods:}  
We searched PubMed (via MEDLINE), Web of Science, and Google Scholar, from inception up to November 2024 to identify articles that propose and evaluate clustering techniques for CTS. Methods were classified according to three major families—distance-based, feature-based, and model-based—and assessed on their ability to handle specific data challenges such as variable sequence length, multivariate data, continuous time, missing data, time-invariant covariates, and large data volumes. \\

\textbf{Results:}  Out of $14607$ studies retrieved from three databases (Web of Science, Pubmed, Google Scholar) through keyword search, we included $124$ articles describing $129$ methods, spanning diverse domains including artificial intelligence, biology, social sciences, and epidemiology. Distance-based methods, particularly those using Optimal Matching, were most prevalent, with $56$ methods. We identified $28$ model-based methods, which demonstrated superior flexibility for handling complex data structures such as multivariate data, continuous time and time-invariant covariates. We also recorded $45$ feature-based approaches, which were on average more scalable but less flexible. Less than half provided publicly available implementations. A searchable Web application \footnote{\url{https://cts-clustering-scoping-review-7sxqj3sameqvmwkvnzfynz.streamlit.app/}} was developed to facilitate method selection based on dataset characteristics. \\


\textbf{Discussion:}  
CTS clustering methods are highly heterogeneous in their assumptions, capabilities, and scalability. While distance-based methods dominate, model-based approaches offer the richest modeling potential but are less scalable. Feature-based methods favor performance over flexibility, with limited support for complex data structures. \\

\textbf{Conclusion:}  
This review highlights methodological diversity and gaps in CTS clustering. The proposed typology and Web application aim to guide researchers in selecting appropriate methods for their specific use cases.

\newpage

\section{Background}

Time series and sequence analysis focus on observations of one or multiple variables structured along an ordered scale, either temporal or non-temporal. These variables can be real-valued, discrete, ordinal, or categorical (i.e., discrete and non-ordinal). Among these, categorical times series or sequences (refereed as CTS throughout this article) represent a specific and challenging case, as they lack a natural metric structure and involve symbolic, non-comparable values. \\

CTS data involve ordered list of categorical observations.They are found in a variety of fields such as epidemiology (e.g., to analyze care trajectories) \cite{6Wmodelcaretrajectories}, in sociology (e.g., to analyze how social sequences can reflect changes in socio-economic variables over time) \cite{SequenceAnalysisPPF}, in genomics (e.g., to analyze DNA sequences) \cite{DNASAbook} or in marketing (e.g., to analyze clickstream data \cite{clickstreammodeling} which represent sequences of clicks on a website). Table $1$ (see 4. Results) illustrates this common structure through examples from the literature. \\

A typical goal in CTS analysis is to find common patterns across individual trajectories. Clustering is often used as a first step in these situations, to help with data visualization, outlier detection, or formulating hypotheses. Clustering methods for CTS have been developed in different fields. In sociology, they are referred to as Sequence Analysis \cite{SequenceAnalysisPPF}. This field started with the work of Abbott et al. \cite{Abbott}, which introduced the Optimal Matching (OM) dissimilarity. OM adapts the Alignment distance and the Needleman–Wunsch algorithm—originally used in biological sequence analysis—to the study of social sequences, with the goal of identifying typologies. \\

Reviews on CTS clustering in the literature tend to be limited in scope and ignore methods coming from other fields \cite{Distancereview, SASystematicReview, Healthscopingreview}. For this reason, we aimed to provide a comprehensive review of CTS clustering methods across multiple disciplines, and a typology of methods based on the type of approach and whether they account for specific cases such as 1) sequences of varying lengths, 2) multidimensional observations, 3) irregular or continuous time sampling, 4) missing data , 5) time-invariant covariates and 6) big volumes of CTS data.

\section{Methods}

This review was reported following the Preferred Reporting Items for Systematic reviews and Meta-Analyses extension for Scoping Reviews (PRISMA-ScR) guidelines \cite{Prisma1}.\\

\textbf{Search Strategy}

We searched on Google Scholar, Web of Science, and PubMed via MEDLINE, from their inception until November 11, 2024. The strategy used as keyword with synonyms of "clustering", "categorical", and "sequence", identified during a preliminary search. To ensure better coverage of domain-specific methods, we also included some expressions, such as "DNA" and "Sequence Analysis", which are frequently used in bioinformatics and social sciences, respectively. The complete list of synonyms and the final search strings for each database are provided in the Appendix. \\

\textbf{Eligibility criteria}

We included articles that presented a method for CTS clustering and tested it on a synthetic or real-life dataset. Studies focusing on temporal numerical, discrete ordinal, or univariate binary sequences were not considered. Articles proposing new dissimilarity measures or probabilistic models for CTS without applying them to clustering, or "clustering by example" approaches that cluster timestamps instead of sequences \cite{HCnominalDatastreams}, were excluded. We also excluded application papers (i.e., articles that only applied a clustering method without describing it in detail within the manuscript text), as well as those without an abstract or not written in English. \\

\textbf{Study Selection}

Retrieved studies were screened using Covidence. Two reviewers (OK and FP) independently screened $100$ randomly selected records by title and abstract, discussing any disagreements. The remaining records were then screened by one reviewer (OK). Next, $142$ studies were screened by two reviewers (OK and FP), and the remainder by one reviewer (OK). \\

\textbf{Data extraction}

In each included article, one reviewer (OK) extracted the CTS clustering methods. If several methods were described in the same paper, we retrieved the original articles describing each method. We also searched all supplementary materials and available information online (e.g., code).\\ 

Furthermore, one reviewer (OK) appraised the dataset used in the article to evaluate each method, characterized by the number of sequences ($N$), the number of observed categorical variables ($m$) ($=1$ in the univariate case), the numbers of levels of each categorical variable $s_1 = |\Sigma_1|, \ldots, s_m = |\Sigma_m|$, and the length of the longest sequence ($T$). When multiple datasets are tested, we retained the one with the largest product $N \times |S_k| \times T$, where $S$ denotes the size of the alphabet of the categorical variable with the highest number of levels. \\

For each method included in the review, we checked whether it was accompanied by publicly available code, package or software. \\

We examined how the number of clusters was determined—whether automatically, user-specified, or selected using post-hoc metrics (e.g., Silhouette score \cite{Silhouette}, or BIC for mixture models \cite{schwarz1978estimating}). We also recorded the number of hyperparameters required, when reported. \\

An additional reviewer (AB) independently checked $10$ randomly selected studies for data extraction.

\newpage

\section{Analysis}

\subsection{Communities and use cases}

Each method was categorized in a given community based on the disciplinary category of the journal provided by Web of Science, or the affiliation of the first author for preprints. We also described the data used in the application of the method and categorized it into eight "use cases" (see Table $1$ in 4.1).

\subsection{Clustering approaches}

Each CTS clustering method has two subtasks. First, it needs a way to treat categorical sequential data. Secondly, it uses a clustering algorithm. The first step has three types of approach, already well defined in the CTS clustering literature \cite{hardandsoft}. They define three families of methods for CTS clustering. These families are : 

\begin{itemize}
    \item \textbf{Distance-based} methods define dissimilarity measures for CTS and typically apply clustering algorithms such as hierarchical clustering or k-medoids to the resulting distance matrices. One notable example is the Optimal Matching (OM) distance, also known as Levenshtein,  sequence alignment or edit distance. In the following, we use the terms 'distance' and 'dissimilarity' interchangeably, even though in many cases these functions do not satisfy the triangle inequality and are sometimes not symmetrical. Any method defining a dissimilarity function on CTS and processing a distance matrix is considered distance-based. 

    \item \textbf{Model-based} methods define generative models (e.g., Markov Chains or Generalized Linear Models) for CTS and estimate their parameters on the dataset. The learned model is then used to compute the clusters. Any method that fits a generative model to the entire dataset and uses it for clustering is considered model-based.

    \item \textbf{Feature-based} methods transform categorical time series into numerical vectors, allowing to process with standard clustering methods. This transformation results in loss of information. Features can be constructed by counting subsequences \cite{kmerpackage}, estimating empirical transition matrices on each sequence \cite{ClickstreamPackage}, or computing autocorrelation vectors \cite{hardandsoft}. Once sequences are transformed in vectors of dimension $d$, processed by generic clustering algorithms like k-means or DBSCAN with any dissimilarity function on $\mathbb{R}^d$. Any method transforming sequences in elements of an vector space is considered feature-based. 
\end{itemize}

We categorized each method according to its family. Within each family, we created subcategories based on the type of distance/featurization/model involved. A comprehensive list with a discussion of each is provided in the appendix.\\ 

Once the distance, model, or featurization approach is set, each method applies a clustering algorithm to group the CTS. We identified the main clustering paradigm used in this second step, such as partition-based (e.g., k-means), hierarchical (agglomerative/divisive), density-based (e.g., DBSCAN), and model-based approaches. 

\subsection{Accounting for specific Data structures}

We checked whether each method 1) could handle sequences of varying lengths 2) handled multivariate CTS (i.e multiple categorical variables observed over time) 3) considered time as a continuous variable rather than discretized; 4) included a mechanism for handling missing values (e.g. imputation); 5) could incorporate time-invariant covariates (e.g., age, gender, biomarkers) in the clustering process; , and 6) could process a big volume of data (defined as  $N \times |S_k| \times T > 10^7$).

\subsection{Dependency order}
We assessed the order of dependency each method can capture. First-order Markov chains, for example, assume each entry depends only on the previous one and thus capture dependencies of order $1$. Hamming distance methods \cite{HammingOriginal}, being insensitive to column order, capture no temporal dependency : order $0$. In contrast, methods based on OM dissimilarity can compare subsequences of any order : their dependency order is set to $\infty$. A detailed explanation of this concept is provided in the Appendix.

\section{Results}

\subsection{Description}

The electronic search retrieved $14607$ unique articles. After the full text review, $124$ articles were included. Among these articles, we identified $129$ methods. Methods were used in eight main communities. The most frequent community was Artificial Intelligence with $30$ ($23.3 \%$) methods, followed by Biology, Social Science (which gathers sociology and economics), Statistics, Computer Science and Healthcare. \\

Methods were tested in a total $171$ applications with data ranging from Biological Sequences (DNA or Protein), to social sequences, care trajectories and clickstream data. A table describing the relationship between methods and communities is given in Appendix.    \\

\begin{figure}[H]
\begin{center} 
\centering
\includegraphics[width = 15cm ]{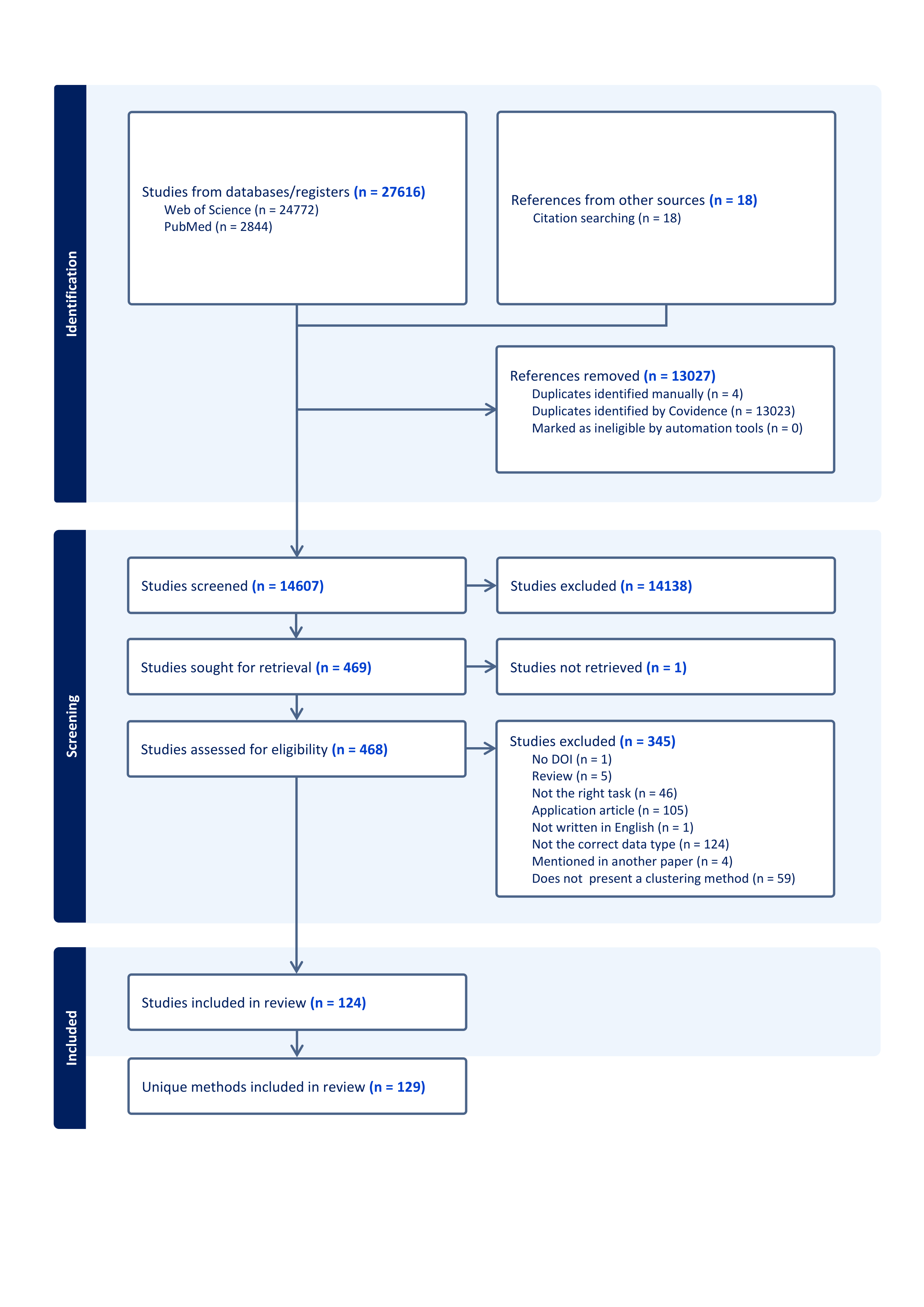}
\caption{Prisma Flow Diagram} 
\end{center}
\end{figure}

\begin{table}[H]
    \centering
    \renewcommand{\arraystretch}{3} 
    \setlength{\tabcolsep}{15pt} 
    \caption{Main use cases for CTS clustering, with number of found occurrences and illustrated examples from the literature}
    \begin{tabular}{|>{\centering\arraybackslash}m{3cm}|c|>{\centering\arraybackslash}m{7cm}|c|}
        \hline
        \textbf{\begin{tabular}[t]{@{}c@{}} Use Case  \end{tabular}} & 
        \textbf{Number} & 
        \textbf{\begin{tabular}[t]{@{}c@{}} Example from the literature \end{tabular}} & 
        \textbf{Ref} \\
        \hline

        \begin{tabular}[t]{@{}c@{}}
            Biological Sequences \\[-10mm] (DNA, Protein)
        \end{tabular}
        & 52 ($30.4 \%$)& 
        \begin{tabular}[t]{@{}c@{}@{}}
            \input{tikzfigures/tikzbiology}\\[-1cm]
            DNA Sequence.
        \end{tabular} & 
        \cite{idelucs} \\
        \hline

         \begin{tabular}[t]{@{}c@{}}
            Social sequences
            
        \end{tabular}
        & 31 ($18.1 \%$)& 
        \begin{tabular}[b]{@{}c@{}}
            
            \input{tikzfigures/tikzsocial} \\[-1cm]
            Bivariate Work and Family Trajectory.
        \end{tabular} & 
        \cite{multichannel} \\
        \hline
        
        \begin{tabular}[t]{@{}c@{}}
            Care trajectories
        \end{tabular}
        & 19 ($11.1 \%$) & 
        \begin{tabular}[t]{@{}c@{}@{}}
            \input{tikzfigures/tikzcare}\\[-1cm]
            Continuous Time Care trajectory for \\[-1cm]rhumatoid artritis patient (source : \cite{reuma}).
        \end{tabular} & 
        \cite{aliclu} \\
        \hline

        \begin{tabular}[t]{@{}c@{}}
            Clickstream Data
        \end{tabular}
        & 14 ($8.2 \%$)& 
        \begin{tabular}[t]{@{}c@{}@{}}
            \input{tikzfigures/tikzclick}\\[-1cm]
            Sequence of user clicks on products (P)\\[-1cm]
            from an online store. 
        \end{tabular} & 
        \cite{ClickstreamPackage} \\
        \hline
        \begin{tabular}[t]{@{}c@{}}
            Activities Sequence
        \end{tabular}
        & 11 ($6.4 \%$) & 
        \begin{tabular}[t]{@{}c@{}@{}}
            \input{tikzfigures/tikzactivity}\\[-1cm] Continuous time sequence
            \\[-1cm] of daily activities.
        \end{tabular} & 
        \cite{HammingMoreau} \\
        \hline
        
  \begin{tabular}[t]{@{}c@{}}
            Speech sequences
        \end{tabular}
        & 6 ($3.5 \%$)& 
        \begin{tabular}[t]{@{}c@{}@{}}
            \input{tikzfigures/tikzspeech}\\[-1cm]
            Sequence of recognized vowels\\[-1cm]
            in a speech.  
        \end{tabular} & 
        \cite{tuplelength} \\
       \hline
     \begin{tabular}[t]{@{}c@{}}
            Clinical trajectories
        \end{tabular}
        & 4 ($2.3 \%$)& 
        \begin{tabular}[t]{@{}c@{}@{}}
            \input{tikzfigures/tikzclinical}\\[-1cm]
            Sequence of multiple clinical categorical \\[-1cm] outcomes
             of a patient. 
        \end{tabular} & 
        \cite{HMMClust} \\
       \hline
        
    \end{tabular}
    
    \label{tab:example1}
\end{table}


Half ($ n = 71, 55.0 \%)$  of the $129$ identified methods did not provide a public implementation of any form. Among the $58$ other methods, we differentiated the proposed methods with a simple publicly available code (typically via a GitHub link) ($n = 30, 23.3\%$) from those implemented in a package ($n = 22, 17.1\%$) or free software ($ n = 6, 4.7\%$). 

\subsection{Identified methods and their characteristics}

Table $2$, $3$ and $4$ summarize all the information we extracted for distance-based, feature-based and model-based methods respectively. Due to space constraints, only the methods provided with public code are presented. Their complete versions are provided in Appendix (section 8.6). \\

\subsubsection{Distance-based methods}

\input{Tableaux/TableauDistances}

\restoregeometry

We identified $n = 56 \ (43.4\%)$ distance-based methods, from which $n=40$ ($71.4\%)$ are based on Optimal Matching (OM) dissimilarity, its variants, or its restrictions such as Hamming distance, Jaro distance, or Longest Common Subsequence (LCS). Several studies, including those cited in \cite{Distancereview}, define new strategies to tune hyperparameters for the OM dissimilarity—especially its substitution costs \cite{OMsloc, OMstran}.\\

Among these $56$ methods, $n = 24$ use hierarchical clustering algorithms ($42.9\%)$, with a marked preference for Ward linkage ($n = 8$). Partitioning algorithms such as $k$-medoids (PAM) are also common with $ n= 9$ methods ($16.1\%)$, and graph-based approaches appear in $n = 6 \ (10.7\%)$ methods, typically through partitioning on neighborhood graphs. A large majority of these methods ($n = 50$, or $89.3\%$) can handle sequences of varying lengths. In particular, OM dissimilarity allows deletion, enabling comparison of sequences of different lengths. However, it is often recommended to pre-process the data to have sequences of the same length (see the chapter of \cite{analyse-R} focusing on Care Trajectories analysis).\\

Multivariate CTS are supported by $n = 9 \ (16.1\%)$ distance-based methods. The most common approach is Multichannel Sequence Analysis (MSA), an extension of OM adapted to multivariate sequences \cite{multichannel}. Another strategy is to use an Extended Alphabet to encode multivariate CTS as univariate sequences. See \cite{Multichannelcomparison} for a comparison of these two approaches. Handling continuous time is possible with $n = 5 \ (8.9\%)$ distance-based methods, using adaptations of OM, Hamming distance, or Dynamic Time Warping \cite{aliclu, HammingMoreau, hierastiseq}.\\

Missing data are explicitly considered by $n = 5 \ (8.9\%)$ distance-based methods. The only documented approach is to consider "Missing" as an additional categorical state, despite its limitations—such as its tendency to generate "artificial" clusters of sequences with missing values, as documented in \cite{missingbenchmark}. Furthermore, only one distance-based approach for handling time-invariant covariates has been recorded \cite{lemeur}, with an application on care trajectories.\\

The computational cost of the OM dissimilarity is high and has been widely addressed in bioinformatics. As a result, classical OM-based methods are generally applied to relatively small datasets, with the product $N \times S \times T$ ranging from $10^5$ to $10^7$. To address scalability, several greedy heuristics have been proposed \cite{blastclust, cdhit, Clover} to avoid computing all pairwise distances, and tested on large-scale biological datasets. For these methods, the product $N \times S \times T$  can exceed $10^{10}$, with a record of $10^{12}$. We refer to them as "selective" distance-based methods.\\

Among distance-based methods, $n = 11 \ (19.6 \%) $ have no reported hyperparameters. The most frequent hyperparameters are weights, with a focus in the literature on the substitution cost schemes (one weight per pair of categorical attributes) of OM, reported in $n = 10 \ (18.0\%)$ methods. These schemes can be expert-driven (e.g., for DNA sequences) or chosen automatically. See \cite{Distancereview} for an overview of automatic (or data-driven) substitution cost scheme strategies. Weights are also needed for mixed Euclidean distances, to control the importance of categorical features \cite{panelgower}. Also, $n = 6 \ (10.7\%)$  distance-based methods use rules requiring the choice of a similarity threshold (e.g., greedy approaches \cite{blastclust} or graph-based algorithms \cite{BAG}).

\subsubsection{Feature-based methods}

\input{Tableaux/Tableaufeature}
\restoregeometry

\newpage

We recorded $n = 45 \ (34.9\%) $ feature-based methods, showing a wide variety in how feature vectors are constructed from CTS. The most common strategy is the use of $k$-mers (or $n$-grams) vectors, found in $n = 13 (28.9\%)$ methods, where sequences are represented by their subsequences of given length. These vectors are either clustered directly \cite{kmerpackage}, \cite{RKMS}, or used in self-supervised learning pipelines \cite{MeshClust}, \cite{Meshclustv3}. Other featurization techniques include Chaos Game Representation \cite{CGR}, empirical transition matrices \cite{ClickstreamPackage}, Fourier transform \cite{EnvClust}, or latent space representations obtained from deep learning architectures \cite{UserJourney}. \\

Once transformed into feature vectors, CTS can be clustered using standard algorithms for numerical data. Among the $45$ recorded approaches, $k$-means and its variants such as k-medoids or bisecting k-means are the most frequently used ($n = 21$ methods, or $46.7\%$), followed by hierarchical clustering, used in $n = 17$ methods $(37.8\%)$. \\

All of the $45$ recorded feature-based methods can handle CTS of various lengths.  However, many of these methods have been developed in bioinformatics and focus on DNA, RNA, or protein sequences. As a result, certain specific data structures remain underexplored. Continuous Time is adressed  in $n = 2$ methods \cite{NMSSVR, CFDApackage} $(4.4\%)$, and $4$ \cite{Transitionmatrixfeatures, UserJourney, MiningClinicam, Typical}, handle multivariate CTS $(8.9\%)$. Only one method adresses the issue of missing data \cite{ClickstreamPackage}. No feature-based method can handle time-invariant covariates. \\

Many of these methods have been developed in bioinformatics to process large datasets without computing all pairwise dissimilarities. "Aside from the 'selective' distance-based methods mentioned earlier, feature-based methods are those that allow for handling the largest data volumes, particularly in the field of bioinformatics. Among them, $n = 8$ (17.8\%) were tested on large-scale datasets, with $N \times S \times T$ ranging from $10^8$ to $10^{11}$.\\

The most common hyperparameter for feature-based approaches encodes the dependency order, as defined in the Appendix, and is reported in $n = 20 \ (44.4\%)$ methods. It may represent the subsequence length $k$ in $k$-mer vectors \cite{kmerpackage}, the size of a sliding window \cite{ntreeclus}, or the order of Markov models \cite{ClickstreamPackage}. This hyperparameter determines both the complexity of dependencies the method can capture and the required computation time. Among feature-based methods, $n=9 \ (20\%)$  rely on ML algorithms and thus require to specify hyperparameters such as learning rate, number of layers for deep learning approaches \cite{UserJourney} or number of trees for Random Forest approaches \cite{ntreeclus}.

\subsubsection{Model-based methods}

\input{Tableaux/Tableaumodel}
\restoregeometry

\newpage

We identified $n = 28$ model-based methods, representing $21.7\%$ of the total. The most widely used models are Markov chains and Hidden Markov Models (HMMs) and their variants, with $n = 16$ methods $(57.1\%)$. Generalized Linear Models (GLMs) have also been adapted for CTS data in $n = 6$ recorded approaches $(21.4\%)$. The Expectation-Maximization (EM) algorithm is used in $n= 17$ methods $(60.7\%)$ to estimate parameters of a mixture model. For Bayesian approaches, estimation is carried out using Monte Carlo Markov Chain algorithms in $n = 7$ methods ($25\%$). \\

Most model-based methods are applicable to CTS of various lengths ($n = 25, 89.3\%$). A subset of $n = 9$ methods extend to multivariate CTS ($32.1\%$), including adaptations of GLMs\cite{GBTMoverview, lamb} and  HMMs \cite{seqHMM, Twostepheterogeneous}. A method based on Transformer architectures \cite{UserJourney} has also been proposed. \\

Continuous time is adressed in $n = 10$ methods ($35.7\%)$. This includes adaptations of Markov chain models \cite{Impactofestimation, Nonstationarymarkov}, time-dependent Generalized linear models \cite{GLMclust, ClusterwiseGLM}, and methods based on functional latent representations for CTS \cite{lookaliker, CFDApackage}. Missing data are explicitly considered in $n = 10$ methods ($35.7\%)$, mostly through imputation using the defined generative model. Time-invariant covariates are incorporated in $n = 11$ model-based approaches ($39.3\%)$, typically as predictors of initial cluster membership probabilities in each cluster via a multinomial logit model \cite{markovbayeslogit}, or as fixed-effect in GLMs \cite{lamb}.  \\

Model-based approaches are notoriously less scalable than other families due to the computational complexity of algorithms such as  EM or MCMC. As a result, only $n = 4$ methods have been tested on large datasets ($14.3\%)$, and most approaches are tested on moderate data volumes, with $N \times S \times T$ ranging from $10^4$ to $10^7$.  \\

All of the $n = 17$ methods relying on the EM algorithm require initialization for this iterative procedure. In practice, this is often done via grid search. Bayesian approaches require the specification of a prior for the parameters of interest. One method based on Markov chains leaves the choice of dependency order open \cite{BayesianClustering}, while all others fix it to $1$. Methods based on GLMs require the user to specify the model structure (e.g., inclusion of polynomial terms in time or the form of random effects), as well as to choose a baseline category to compute probability ratios, for identifiability reasons.

\subsection{Number of Clusters choice}

Among the $129$ identified methods, $n = 55$ did not provide a generic way to choose the number of clusters $(42.6\%)$, leaving the decision to the user. Among the remaining methods, $n = 19 \ (14.7\%)$ implemented clustering algorithms that automatically determine the number of clusters (for example, through an explicit stopping rule). The others rely on "post-hoc" criteria to compare multiple clustering solutions. The number of clusters is selected using AIC or BIC (Akaike Information Criterion and Bayesian Information Criterion, respectively), in $n = 18$ methods ($14.0\%)$. These criteria are well suited for model-based approaches as they balance model likelihood and complexity. Among distance-based and feature-based methods, the most popular family of criteria are those based on the Silhouette score \cite{Silhouette}, such as the Average Silhouette Width (ASW), used in $n = 13$ methods. Many other criteria are mentioned, but no other trend stands out as clearly. \\

\subsection{Web Application}

To make navigation through our results as easy as possible, we built a Streamlit Web Application \footnote{\url{https://cts-clustering-scoping-review-7sxqj3sameqvmwkvnzfynz.streamlit.app/}}. This application allows to filter methods based on a combination of specific cases (e.g : covariates and multivariate) or on tested use-cases (e.g if the user only wants methods that have been already tested on care trajectories). Its code and the list of methods are open source and can be improved based on readers suggestions.

\section{Discussion}

Our scoping review identified $129$ methods for CTS clustering. The variety of approaches stems from two main challenges in CTS analysis: giving a metric structure to categorical data, and modeling dependency. From these two challenges, three types of approaches emerged from the literature: distance-based, feature-based and model-based CTS clustering, each with several subcategories. 

\subsection{Choosing a method based on dataset characteristics}

To pick the right CTS clustering method, researchers should account for six specific data characteristics: sequences of various lengths, multivariate data, continuous time, missing data, the presence of time-invariant covariates, and big volumes of data. These issues are key in epidemiology, where irregular sampling, missing data, censoring, and the inclusion of time-invariant covariates are frequent. \\

The issue of sequences of various lengths is the most commonly addressed: $n = 118$ methods are able to deal with it ($91.5\%)$. Popular approaches such as Optimal Matching, $k$-mers, and Markov chains all support variable-length sequences. 
The multivariate case is less covered, but still considered in $n = 23$ studies ($17.8\%)$. Continuous time is more rarely addressed: we identified $n= 18$ methods explicitly capable of handling it ($14.0\%)$ \\

Missing data is considered in $n = 16$ included studies $(12.4\%)$, but remains overlooked in the literature. This question deserves a review or benchmark on its own, and goes beyond the question of clustering. Recent work in this direction has been conducted by  Emery et al. \cite{emery2024comparison} in the univariate and multivariate cases. In principle, any model-based approach could be used for missing data imputation, and some explicitly implement it \cite{seqHMM}. \\

A few methods ($n = 12, 9.3\%$) are able to handle time-invariant covariates in the clustering procedure. Notably, model-based clustering methods are — with the exception of \cite{lemeur} — the only class of methods that can include covariates directly. Standard approaches include estimating initial cluster probabilities using a multinomial logit model \cite{MultinomialLogit}, \cite{seqHMM}, \cite{GLMclust}, or including covariates as fixed effects in GLM-based models \cite{lamb}, \cite{ClusterwiseGLM}. \\

In our review $n = 37$ ($28.7\%)$ methods had been tested on large volumes of data as defined in section 3.2. Most of these were developed in the bioinformatics field, often relying on feature-based or selective distance-based strategies to avoid computing all pairwise dissimilarities. \\

Almost half ($n = 61, \ 47.3\%$) of the identified methods were not provided with public code, package or software. Although these methods are typically described through pseudocode or in the main text, this lack of implementation hinders both the reproducibility of the methods and their associated experiments. It also creates a significant time burden for researchers who need to reimplement these methods for their own work. 

\subsection{A comparison of the three method families for CTS clustering}

Distance-based approaches remain the most common, with OM being the most popular approach overall, with many variants developed for sociology. Extensions of OM have been proposed to address specific cases such as multivariate data \cite{multichannel} or continuous time \cite{aliclu} , but none allows for handling all the considered specific cases simultaneously. OM also raises critiques regarding the choice of its hyperparameters — especially substitution costs \cite{Distancereview} — its lack of scalability, and its tendency to bias clustering results when sequences of various lengths or missing data are present \cite{analyse-R, missingbenchmark}. Its popularity is nevertheless explained by its great flexibility, interpretability, the absence of underlying assumptions, and the availability of reliable software to use it \cite{traminer}. \\

Among the three families, model-based methods offer the best ability to handle specific cases. Methods based on time-dependent Generalized Linear Models are particularly capable, and \cite{ClusterwiseGLM} even addresses all five considered problems simultaneously (though no available code is provided with it). However, these approaches are also the most demanding: they are harder to use, involve more hyperparameters (e.g. prior choice for Bayesian approaches, choice of a baseline categorical state for GLMs, EM initialization), and rely on stronger assumptions about the generative process of the data. They also scale less easily, although some methods have been tested on moderately large datasets \cite{biclustering}, \cite{LinearMixedEffect}, \cite{ClusterwiseGLM}. \\

At the other end, feature-based approaches remain the least able to handle specific structural cases, especially when they appear simultaneously, due to their focus on biological sequences. They present a large variability in how feature vectors are constructed and processed, with $k$-mers being the most popular approach. Many of them were developed to avoid computing all alignment distances; for this reason, they are often referred to as "alignment-free" methods \cite{alignmentfreereview}, especially in large-scale applications. A feature-based approach that could simultaneously adress several of the specific cases we focused on while remaining scalable seems to us a promising direction for future research.

\subsection{Limitations}

Our review has some limitations. First, even though the number of identified methods ($129$) can be considered high, we had to make choices in the keyword search that may have caused us to miss a few, or even entire fields of application for CTS clustering. Second, it does not seem reasonable to benchmark all of the $129$ methods at a time, since we showed that they are not designed exactly for the same use-cases.  Moreover, even benchmarking a few methods inside a specific use-case requires choosing one or several generative mechanisms to build synthetic datasets and assessing each clustering method knowing the ground truth, and one or several real-life datasets, for example to perform a sensitivity analysis. This falls outside the scope of this review, but could constitute another work in its own right. Thirdly, as we focused exclusively on the clustering task, we overlooked many CTS analysis tools and models, since they did not provide clusters natively.

\section{Conclusion}

This scoping review has highlighted the multidisciplinarity and variety of approaches to the CTS clustering problem. We provided a comprehensive methodological review with $129$ included methods. We proposed a Web Application to choose the right CTS clustering method in regard of the data characteristics. 

\section{Contributors}

OK, FP and VTT conceptualized this scoping review. OK and FP screened the articles. OK, FP and AB participated in the full-text data extraction. OK and VTT drafted the manuscript. All authors participated in revisions of the manuscript. 
\newpage

\section{Appendix}

\subsection{A formal definition for CTS}

\begin{definition}
\normalfont
    
Throughout the article, we will call a \textbf{categorical} variable any variable that can take a finite number of values that can not be ordered.
\end{definition}
This definition contrasts with the notion of \textbf{ordinal} variable, when the categories can be ordered. For example, it excludes count data. It is similar to the definition that \cite{agresti2012categorical} gives of a \textbf{nominal} scale variable. The set of categories will be denoted $\Sigma = \{ s_1 \ldots s_c\}$, which can be interpreted as an alphabet. We define a categorical time series as follow. 

\begin{definition}
\normalfont
    A multivariate time series with categorical features and covariates is the data of a pair $Z = (X,Y)$ where : 
    \begin{itemize}
        \item[$\bullet$] $X = (X_1 \ldots X_k)^T$ is a vector of time-invariant covariates, which can be continuous, ordinal or categorical. 
        \item[$\bullet$] $Y = ((Y_1 , t_1) \ldots (Y_T , t_T))$ is an ordered sequence of random vectors $Y_i$ associated with times $t_i \in \R$ with $t_1 < \ldots < t_T$. 
        \item[$\bullet$] $Y_i = (Y_i^1, \ldots Y_i^{m_1}, Y_i^{m_1 + 1} \ldots Y_i^{m_1 + m_2})^T $ is the $i$-th observation, consisting of $m_1$ categorical variables $Y^i_1 \ldots Y^i_{m_1} $, each of which can take its values from alphabets $\Sigma_1 \ldots \Sigma_{m1}$, and $m_2$ additional features $Y^i_{m_1 + 1} \ldots Y^i_{m_2}$ which can be continuous or ordinal. 
    \end{itemize}
\end{definition}

Most of the methods seen through the review are limited to the case of $k = 0$ (no covariates), $m_1 = 1$ and $m_2 = 0$ (univariate categorical sequences) and assume perfectly regular sampling, which can be summarized by the assumption $t_1 = 1, \ldots, t_T = T$. 

\subsection{Dependency order definition}

In the following, we will use the previous definition of a categorical time series dataset $Y$, with regular sampling $t_1 = 1 , \ldots , t_T = T$. For simplicity, we will focus on the univariate case, although all the following concepts can be generalized to multivariate data. Our dataset is $Y_{[n]} = (Y^j)_{1 \le j \le N}$ where $Y^j = (Y_1^j, \ldots Y_{T_j}^j)$ is an univariate categorical time series with values in an alphabet $\Sigma$ and of length $T_j$. For the sake of simplicity, we assume all sequences have the same length. Everything that follows can be generalized to sequences from various lengths by completing sequences with symbols $? \notin \Sigma$. \\

Each $Y_i^j$ is a random variable with values in $\Sigma$, and the time series $Y^j$ are assumed to be i.i.d. observations from an unknown underlying distribution.  $Y_{[n]}$ lies in the space $\mathcal{Y}$ of all possible categorical time series datasets of any size, defined as :

$$\bigsqcup_{n \in \N} \left(\bigsqcup_{T \in \N} \Sigma^T \right)^n$$

In this context, a clustering on the dataset $Y_{[n]}$ can be viewed as an element of the set $\mathcal{B}(N)$ of partitions of $\llbracket 1 ; N \rrbracket$. It depends on $Y_{[n]}$ and other exogeneous random variables $Z$, such as the random initialization of a $k$-means procedure or EM algorithm, lying in a space $\mathcal{Z}$. 

\begin{definition}
    A \textbf{clustering method} $C$  is a function : 

\begin{align*}
C \colon &~ (\mathcal{Y}, \mathcal{Z}) \to \mathcal{B}(N) \\
&~ (Y_{[n]}, Z) \mapsto C(Y_{[n]}, Z)
\end{align*}

\end{definition}

A column of $Y_{[n]}$ is denoted $Y_i$ for any $i \in \llbracket 1 ; T \rrbracket$. $Y_i$ is a vector over $\Sigma$, of length $n$. 

\begin{definition}
    The set $G_k\left(Y_{[n]}\right)$ of \textbf{$k$-grams} of $Y_{[n]} \in \mathcal{Y}$ is the set of all possible $k$-uples of consecutive columns in $Y_{[n]}$, i.e : 

    $$G_k\left(Y_{[n]}\right) = \left\{ (Y_i \ldots Y_{i+k} ), i \in \llbracket 0 ; T-k \rrbracket\right\}$$
\end{definition}

Denoting $\Sigma^* = \bigsqcup_{n \in \N} \Sigma^n$, $G_k(Y_{[n]})$ lies in the space : 

$$\mathcal{G} = \bigsqcup_{k \in \N} \bigsqcup_{T \in \N} \left( \left( \Sigma^* \right)^k \right)^T$$

and $G_k$ is a function from $\mathcal{Y}$ to $\mathcal{G}$. 

\begin{rmk}
    $G_k(Y_{[n]})$ is the set of $k$-uples of consecutive columns in $Y_{[n]}$, considered without order. In particular, $G_1(Y_{[n]})$ is the set of columns of $Y_{[n]}$, regardless of their order. 
\end{rmk}

\begin{definition}
 Let $C : (\mathcal{Y}, \mathcal{Z}) \to \mathcal{B}(N)$  be a clustering method. $C$ is said to have a \textbf{dependency order} of at most $k$ if there exists $k$ such that there exists a function $C_{k+1} : \mathcal{G} \times \mathcal{Z} \to \mathcal{B}{[n]}$, such that : 

 $$\forall Y_{[n]} \in \mathcal{Y},  C(Y_{[n]}, Z) = C_{k+1} \left( G_{k+1}\left(Y_{[n]}\right), Z \right)$$

The minimal $k$ is called the dependency order of $C$.  If such $k$ does not exist, we set it to $\infty$. 
\end{definition}

In other words, a clustering algorithm $C$ takes into account dependencies of order $k$ at most if it can be entirely built from $k+1$-Grams, i.e the unordered set of $k+1$-uples of consecutive columns in $Y_{[n]}$. \\

For example, clustering methods based on first-order Markov models \cite{bayesmarkov} use transition probabilities, which can be entirely computed from $2$-Grams (i.e pairs of consecutive columns), hence have a dependency order of $1$. Clustering methods independent to the order of columns (i.e timestamps) such as Hamming distance \cite{hamming1950error} or Group-based trajectory modelling \cite{GBTMoverview} can be built from $1$-Grams, hence have a dependency order of $0$. Similarly, feature-based methods using $k$-mers vectors have a dependency order of $k$, controlled by the user \cite{kmerpackage}. \\ 

In contrast, methods based on Optimal Matching (OM) \cite{Abbott} can align subsequences from arbitrary length, hence have a dependency order of $\infty$. The same applies to Generalized Linear Models incorporating time as a feature \cite{GLMclust}, or to feature-based methods based on the Fourier transform \cite{FourierTransform}.

\subsection{Keyword list and search equation}

\begin{table}[ht]
\centering
\caption{List of identified synonyms for "Clustering" (\textbf{\#1}), "Categorical" (\textbf{\#2}) and "Sequences" (\textbf{\#3})}
\begin{tabular}{|>{\RaggedRight\arraybackslash}p{3cm}|>{\RaggedRight\arraybackslash}p{9cm}|}
\hline
\textbf{Concept} & \textbf{Synonyms and related terms} \\ \hline
\textbf{\#1 Clustering} & Cluster* OR Partition* OR Dissimilarit* OR Optimal Matching OR Mixture* OR Group* OR Latent Class* \\ \hline
\textbf{\#2 Categorical} & Categorical OR Nominal OR Binary OR Mixed OR Qualitative OR Event OR Discrete OR State* OR Life-course OR Care OR Clinical \\ \hline
\textbf{\#3 Sequences} & Sequence* OR Time Series OR Temporal OR Sequential OR Longitudinal OR Time-evolving OR Time-dependent OR Data Stream* OR Pathway* OR Trajector* OR Repeated OR Panel \\ \hline
\end{tabular}
\end{table}

Some of these terms are not actual synonyms for the intended concept, but they are of interest in a specific community. For instance, "Optimal Matching" is a widely-used name for a distance between categorical sequences, also known as the Levenstein or the Sequence Alignment distance, often used as a dissimilarity for clustering algorithms such as Ward or k-medoids. This expression is especially used in sociology, such as in the review \cite{Distancereview}. Another example is the expression "care trajectory" (or "clinical pathway"), which is a frequent use-case for categorical sequence analysis in the epidemiological field. \\

\textbf{Domain-specific expressions}


In addition to these keywords, we identified two domain-specific expressions used in some communities. 

\begin{itemize}
    \item The first one is "Sequence Analysis." This generic term is especially used in the sociology community to describe the analysis— including clustering—of categorical time series that represent the evolution of a social characteristic for an individual, such as socioprofessional category. For a narrative review of this field in quantitative sociology, see \cite{SequenceAnalysisPPF}.

    \item The second term is "DNA." DNA sequences are a specific type of categorical sequence data with four categories, corresponding to each nucleotide type (A, T, C, and G). The foundational article of the "Sequence Analysis" field in sociology \cite{Abbott} itself is based on approaches that were already well-established and widely used in the bioinformatics field. DNA sequence clustering, for instance, is a standard task for exploratory data analysis, mutation detection, and constructing phylogenetic trees using hierarchical algorithms such as UPGMA. For an empirical review of popular tools in this domain, see \cite{bioinforeview}. Given the multidisciplinary scope of this review, we found it essential to include some sequence clustering methods from bioinformatics.
\end{itemize}

The keywords "Sequence Analysis", "DNA", "Group" or "Latent Class" being very unspecific and present in a very large number of unrelated articles, we have chosen to search them only in title. With the expressions already given by \#$1$, \#$2$ and \#$3$ and with \#$1$' denoting \#1 after removing "Group" and "Latent Class", the exact search equation used on Web of Science is : 

\begin{center}
\begin{align*}
    & (\text{TITLE} = \#1 \textbf{ AND } \#2 \textbf{ AND } \#3) \\
    \textbf{OR } & (\text{ABSTRACT } =  \#1'  \textbf{ AND } \#2 \textbf{ NEAR\textbackslash 1 } \#3) \\
     \textbf{OR } & (\text{TITLE} = \#1 \textbf{ AND } \text{"DNA"})\\
     \textbf{OR } & (\text{TITLE} = \text{"Sequence Analysis"} \textbf{ AND }\text{ABSTRACT } = \#1) 
\end{align*}
    
\end{center}

\subsection{Table of Communities}
\begin{table}[H]
    \centering
    \begin{tabular}{|l|c|p{6cm}|l|}
        \toprule
        \textbf{Community} & \textbf{Number of Methods} & \textbf{Main Use Case} & \textbf{Example} \\
        \midrule
        Healthcare & 11 (8.5\%) & Care Trajectories & \cite{aliclu} \\
         \hline
        Social Science & 16 (12.4\%) & Social Sequences & \cite{DivisiveSocial} \\
         \hline
        Biology & 29 (22.5\%) & DNA Sequences & \cite{idelucs} \\
         \hline
        Statistics & 15 (11.6\%) & Social Sequences & \cite{seqHMM} \\
         \hline
        Artificial Intelligence & 30 (23.2\%) & Clickstream data & \cite{RandomizedAlgorithm} \\
         \hline
        Computer Science & 15 (11.6\%) & DNA Sequences & \cite{LLCS}\\
         \hline
        Engineering & 9 (7.0\%)& Various & \cite{TDACTS} \\
         \hline
        Mathematics & 2 (1.5\%)& Various & \cite{CFDApackage} \\
         \hline
        Misc & 2 (1.5\%)& DNA Sequences & \cite{BORODOVSKY1994259} \\
   
        \bottomrule
    \end{tabular}
    \caption{Repartition of communities for the $129$ listed methods.}
    \label{tab:example2}
\end{table}

\subsection{Definition of method subfamilies}

\begin{itemize}
    \item \textbf{Hamming} : Hamming distance is the simplest CTS dissimilarity, simply counting the number of positions at which states of equal length sequences are different. \\ 
    \item \textbf{OM (Optimal Matching)} : also known as the Levenstein distance, the Alignment distance, or the Edit distance. Allows three operations to transform a sequence into another : substitutions, insertions, and deletions, and uses dynamical programming (most often, the Needleman-Wunsch algorithm \cite{needleman1970general}) to find an optimal sequence of such edit operations with respect to a certain cost scheme. This dissimilarity measure has many variants and special cases, which mostly depend on the operations allowed and how their costs are defined. \\
    \item \textbf{Selective OM} : The computation of all pairwise OM dissimilarities is notoriously intensive, so some methods use heuristics to avoid calculating all of them, replacing most computations with simpler approximations and computing the actual OM distance only when certain conditions are met. \\
    \item \textbf{Selective Hamming} In particular, \cite{Clover} uses a "selective" version of the Hamming distance to cluster a huge DNA sequence dataset. \\
    \item \textbf{LCS (Longest Common Subsequence)} : a particular case of the OM dissimilarity allowing only insertion and deletion, without substitution. This is equivalent to computing the length of the longest common subsequence. \\
    \item \textbf{BLAST} : a similarity score built on $k$-mers heuristics to quickly find local alignments. Can be viewed as a faster approximation of a local variant of OM. \\
    \item \textbf{Correlation} : two identified methods \cite{garcia2015framework, hardandsoft}  define correlation measures on CTS and use it, respectively to define a dissimilarity measure, and feature vectors based on autocorrelations. \\
    \item \textbf{Jaccard} : The Jaccard distance is a dissimilarity measure between two generic sets, calculated as the ratio of the size of their symmetric difference to the size of their union. It can be applied to CTS comparison, taking for instance the $k$-mers as sets to be compared. \\
    \item \textbf{Random Search} : two recorded approaches \cite{RandomizedAlgo, slymfast} use random subsequences to detect patterns in order to build non-standard similarity scores. \\
    \item \textbf{Self-supervised learning} : some methods use self-supervised learning via Machine Learning models  on CTS (forecasting \cite{ntreeclus} \cite{UserJourney}, classification between true and false "decoy" sequences \cite{RFSC}) to build dissimilarity measures or feature vectors and perform clustering. \\
    \item \textbf{KL (Kullback-Leibler)} : the Kullback-Leibler divergence, also known as relative entropy, is a non-symetric dissimilarity measure between probability distributions. It can be used for CTS clustering, associating one probability distribution (using HMM \cite{HMMClust} Markov \cite{Agnostic} or Markov models for instance) at each sequence.  \\
    \item \textbf{Rank} : defined in \cite{Rank} for DNA sequences, the rank distance consists in first transforming each sequence into a string of uniquely annotated symbols, then computing the sum of absolute differences between the positions (or "ranks") of each corresponding symbol in the two sequences. \\
    \item \textbf{Dynamic Time Warping} : dynamic time warping is a dissimilarity measure for time series (generalizable to CTS) that finds an optimal match, allowing non-linear warping in time. It therefore naturally applies to the continuous time case. \\
    \item \textbf{Markov} : Markov chains are a natural choice for CTS modeling, based on the simple assumption that the state at time $t$ depends only on the past states up to a certain order. In the homogeneous case, a Markov chain is therefore described by one matrix containing all transition probabilities. They can be used directly in a mixture model setting for model-based clustering \cite{bayesmarkov}, or as a heuristic to build dissimilarity measures \cite{Fuzzyintegral} or feature vectors \cite{ClickstreamPackage}. \\
    \item \textbf{$k$-mers} : $k$-mers (also known as $n$-grams) is a method to build feature vectors from CTS consisting in counting the number of appearences of each possible subsequence of length $k$, where $k$ is an hyperparameter. These feature vectors can then be used as inputs for generic clustering algorithms (e.g : divisive kmeans in \cite{kmerpackage}) or for more complex pipelines (e.g : \cite{Meshclustv3} uses $k$-mers vectors to train a GLM regressor to predict similarity scores in a self-supervised learning setting). \\
    \item \textbf{Frequencies} : some algorithms only rely on computing frequency of each state in each sequence and using them as feature vectors. This is equivalent to calculating $k$-mers vectors for $k = 1$. \\
    \item \textbf{Fourier transform} : As a fundamental tool in signal processing and time series analysis, the Fourier transform and its variants have been adapted to the CTS case to build feature vectors, and used in several clustering methods. \\
    \item \textbf{Chaos Game Representation} : Chaos Game Representation (CGR) is a method for transforming categorical sequences into images, developed for DNA sequence analysis. These images can then be used as feature vectors in complex sequence clustering pipelines such as \cite{DeLUCS}.  \\
    \item \textbf{Functional data analysis} : the method presented in \cite{CFDApackage} adapts the functional data analysis paradigm - which considers each observation (e.g : time series) as a function, to the CTS case. It defines optimal encoding functions for each state of the alphabet and uses them as feature vectors to perform Ward hierarchical clustering. \\
    \item \textbf{SVR (Subsequence Vector Representation)} : proposed in \cite{NMSSVR}, SVR is a general framework to build feature vectors from subsequences counts. It can handle continuous time and a cost matrix scheme for states. It can be viewed as a generalization of $k$-mers. \\
    \item \textbf{Wavelets} : wavelets are a special class of functions that can be used to decompose time series data into different frequency components. \cite{Waveletbased} adapts it to the DNA sequence case to build feature vectors, using an arbitrary scaling on the alphabet.  \\
    \item \textbf{Moments} : the methods presented in \cite{FeatureMoments} defines $12$-dimensional feature vectors for DNA sequences, from nucleotide counts, mean distance between nucleotide repetitions, and variance of these distances. We call it "moments" (from order $0$, $1$ and $2$)-based feature vectors. \\
    \item \textbf{Fractal dimension} : the method presented in \cite{FractalDimension} adapts the generic concept of fractal dimension (often used for time series analysis) to DNA sequence analysis, using arbitrary numerical representation of states, to build feature vectors. \\
    \item \textbf{Latent Dirichlet Model} : the method presented in \cite{PatientTraces} defines a probabilistic model for care trajectories based on latent topics following Dirichlet laws. Model parameters are estimated with Gibbs sampling and the topic assignation probabilites are used to define feature vectors. \\
    \item \textbf{HMM (Hidden Markov Models)} : Hidden Markov Models are a generalization of Markov chains. Observed states are assumed to be generated by a Markov chain with hidden states. They are described by transition probabilities for the hidden Markov chain and emission probabilities from hidden states to observations. They can be used in the mixture model framework for model-based clustering \cite{seqHMM} or as heuristics to build dissimilarity measures \cite{HMMClust} or feature vectors \cite{Multiview}. \\
    \item \textbf{Entropy} : the methods presented in \cite{Animproved, mBKM} both use Shannon entropies on $k$-mers vectors as feature vectors for DNA sequences. \\
    \item \textbf{Exponential} : the method presented in \cite{Exponential} defines an exponential model for CTS based on the Hamming distance and implements it in a mixture model framework. \\
    \item \textbf{GLM (Generalized Linear Models)} : Generalized Linear Models are a standard and flexible approach in statistics. They generalize linear regression models by introducing a link function. In the case of categorical data, the usual link function is the multinomial logit.  Time can be used as a predictor to modelize temporal dependency \cite{GLMclust, lamb}. Group-based trajectory modelling approaches as defined in \cite{GBTMoverview} can also be seen as a particular case of GLM. \\
    \item \textbf{RNN (Recurrent Neural Networks)} : Recurrent Neural Networks are a neural architecture designed to model sequential data by capturing dependencies between successive elements. \cite{lookaliker} and \cite{PolicyMixture} both use RNN structures as generative models and use them in a mixture model framework. \\
    \item \textbf{Combinatorial} : the methods presented in \cite{TESS} and \cite{CombinatorialModel} both use the combinatorial structure of CTS data to build generative models. These models are not defined at the level of individual sequences, but rather for the entire dataset. They are therefore examples of model-based clustering that do not rely on mixtures. \\
    \item \textbf{Hawkes} : Hawkes processes are point processes that model the occurrence of events over continuous time. The method presented in \cite{DMHP} represents CTS data as multivariate count processes (counting the number of occurrences of each state through time) and learns a mixture model based on Hawkes processes. 
    
\end{itemize}

\subsection{Complete table of Methods}

\newpage
\input{Tableaux/TableauDistancesAll}

\newpage
\input{Tableaux/TableauFeatureAll}
\newpage 
\input{Tableaux/Tableaumodelall}
\newpage 
\bibliographystyle{unsrt}
\bibliography{bibliographie.bib}

\end{document}

%% file: tikzfigures/tikzbiology.tex
\begin{tikzpicture}[
    node distance=0.2cm,
    every node/.style={
        draw,                    
        rectangle,               
        fill=blue!10,            
        align=center,
        font=\tiny,
        inner sep=3pt,
        outer sep=0pt,           
        minimum width=0.4cm,     
        minimum height=0.1cm     
    },
    arrow/.style={-Stealth, thick}
]
\node (A) at (-1.0,0) {A};
\node (T) at (-0.2,0) {T};
\node (C) at (0.6,0) {C};
\node (G1) at (1.4,0) {G};
\node (G2) at (2.2,0) {G};
\node (C2) at (3.0,0) {C};
\node (T1) at (3.8,0) {T};
\node (T2) at (4.6,0) {T};
\node (A2) at (5.4,0) {A};

\foreach \i/\j in {A/T, T/C, C/G1, G1/G2, G2/C2, C2/T1, T1/T2, T2/A2} {
    \draw[arrow] (\i) -- (\j);
}

\end{tikzpicture}

%% file: tikzfigures/tikzsocial.tex
\begin{tikzpicture}[
    node distance=0.2cm,
    every node/.style={
        draw,                    
        rectangle,               
        fill=blue!10,            
        align=center,
        font=\tiny,
        inner sep=3pt,
        outer sep=0pt,           
        minimum width=1.8cm,     
        minimum height=0.8cm     
    },
    arrow/.style={-Stealth, thick}
]
\node (A) at (0,0) {Living w/\\parents};
\node (B) at (2.2,0) {Living w/\\partner};
\node (C) at (4.4,0) {Living w/\\partner + child};

\node (D) at (0,-1.1) {Unemployed};
\node (E) at (2.2,-1.1) {Work};
\node (F) at (4.4,-1.1) {Unemployed};

\foreach \i/\j in {A/B, B/C, D/E, E/F} {
    \draw[arrow] (\i) -- (\j);
}

\draw[thick, decoration={brace, amplitude=5pt}, decorate] (-0.9,-1.6) -- (-0.9,0.5);
\end{tikzpicture}

%% file: tikzfigures/tikzcare.tex
\begin{tikzpicture}[
    node distance=0.4cm,
    every node/.style={
        draw,                    
        rectangle,               
        fill=blue!10,            
        align=center,
        font=\tiny,
        inner sep=3pt,
        outer sep=0pt,           
        minimum width=0.4cm,     
        minimum height=0.1cm     
    },
    arrow/.style={-Stealth, thick}
]
\node (A) at (-1,0) {Etanercept \\
580 days};
\node (B) at (1,0) { Tocilizumab \\ 408 days};
\node (C) at (3,0) { Abatacept \\ 211 days};
\node (D) at (5,0) {End of \\ treatment};

\foreach \i/\j in {A/B, B/C, C/D} {
    \draw[arrow] (\i) -- (\j);
}

\end{tikzpicture}

%% file: tikzfigures/tikzclick.tex
\begin{tikzpicture}[
    node distance=0.2cm,
    every node/.style={
        draw,                    
        rectangle,               
        fill=blue!10,            
        align=center,
        font=\tiny,
        inner sep=3pt,
        outer sep=0pt,           
        minimum width=0.4cm,     
        minimum height=0.1cm     
    },
    arrow/.style={-Stealth, thick}
]
\node (A) at (-1.0,0) {P5};
\node (T) at (-0.2,0) {P1};
\node (C) at (0.6,0) {P6};
\node (G1) at (1.4,0) {P7};
\node (G2) at (2.2,0) {P6};
\node (C2) at (3.0,0) {P7};
\node (T1) at (3.8,0) {P8};
\node (T2) at (4.6,0) {P7};
\node (A2) at (5.4,0) {Buy};

\foreach \i/\j in {A/T, T/C, C/G1, G1/G2, G2/C2, C2/T1, T1/T2, T2/A2} {
    \draw[arrow] (\i) -- (\j);
}

\end{tikzpicture}

%% file: tikzfigures/tikzactivity.tex
\begin{tikzpicture}[
    node distance=0.2cm,
    every node/.style={
        draw,                    
        rectangle,               
        fill=blue!10,            
        align=center,
        font=\tiny,
        inner sep=3pt,
        outer sep=0pt,           
        minimum width=0.4cm,     
        minimum height=0.1cm     
    },
    arrow/.style={-Stealth, thick}
]
\node (A) at (-1.0,0) {Home \\ 330mn};
\node (B) at (0.2,0) {Work\\ 240mn};
\node (C) at (1.4,0) {Lunch \\ 90mn};
\node (D) at (2.6,0) {Walk \\ 60mn};
\node (E) at (3.8,0) {Work \\ 180mn};
\node (F) at (5,0) {Home \\ 630mn};

\foreach \i/\j in {A/B, B/C, C/D, D/E, E/F} {
    \draw[arrow] (\i) -- (\j);
}

\end{tikzpicture}

%% file: tikzfigures/tikzspeech.tex
\begin{tikzpicture}[
    node distance=0.2cm,
    every node/.style={
        draw,                    
        rectangle,               
        fill=blue!10,            
        align=center,
        font=\tiny,
        inner sep=3pt,
        outer sep=0pt,           
        minimum width=0.4cm,     
        minimum height=0.1cm     
    },
    arrow/.style={-Stealth, thick}
]
\node (A) at (-1.0,0) {"A"};
\node (T) at (-0.1,0) {"E"};
\node (C) at (0.8,0) {"A"};
\node (G1) at (1.7,0) {"I"};
\node (G2) at (2.6,0) {"E"};
\node (C2) at (3.5,0) {"O"};
\node (T1) at (4.4,0) {"U"};
\node (T2) at (5.3,0) {"E"};

\foreach \i/\j in {A/T, T/C, C/G1, G1/G2, G2/C2, C2/T1, T1/T2} {
    \draw[arrow] (\i) -- (\j);
}

\end{tikzpicture}

%% file: tikzfigures/tikzclinical.tex
\begin{tikzpicture}[
    node distance=0.2cm,
    every node/.style={
        draw,                    
        rectangle,               
        fill=blue!10,            
        align=center,
        font=\tiny,
        inner sep=3pt,
        outer sep=0pt,           
        minimum width=1.2cm,     
        minimum height=0.8cm     
    },
    arrow/.style={-Stealth, thick}
]
\node (A) at (0,0) {Smoking};
\node (B) at (1.7,0) {Smoking};
\node (C) at (3.4, 0) {Quit \\ smoking};

\node (D) at (0,-1.1) {No CHD};
\node (E) at (1.7,-1.1) {No CHD};
\node (F) at (3.4,-1.1) {No CHD};

\node (G) at (0,-2.2) {Over\\weight};
\node (H) at (1.7,-2.2) {Over\\weight};
\node (I) at (3.4,-2.2) {Obese};

\node (Col4_A) at (5.1,0) {Smoking};
\node (Col4_B) at (5.1,-1.1) {CHD};
\node (Col4_C) at (5.1,-2.2) {Obese};

\foreach \i/\j in {A/B, B/C, D/E, E/F, G/H, H/I}  {
    \draw[arrow] (\i) -- (\j);
}

\draw[arrow] (C) -- (Col4_A); 
\draw[arrow] (F) -- (Col4_B); 
\draw[arrow] (I) -- (Col4_C); 

\foreach \i/\j in {A/B, B/C, D/E, E/F, G/H, H/I}  {
    \draw[arrow] (\i) -- (\j);
}

\draw[thick, decoration={brace, amplitude=5pt}, decorate] (-0.7,-2.7) -- (-0.7,0.5);
\end{tikzpicture}

%% file: Tableaux/TableauDistances.tex
\newpage
\newgeometry{top=2cm, bottom=2cm, left=0.5cm, right=0.5cm}

\begingroup
\scriptsize
\setlength{\extrarowheight}{1.15mm}

\begin{longtable}[c]{|
    >{\columncolor{gray}\centering\arraybackslash}m{1.8cm}|
    >{\columncolor{cblue}\centering\arraybackslash}m{1.8cm}|
    >{\columncolor{gray}\centering\arraybackslash}m{1.8cm}|
    >{\columncolor{cblue}\centering\arraybackslash}m{1.7cm}|
    >{\columncolor{gray}\centering\arraybackslash}m{1.3cm}|
    >{\columncolor{cblue}\centering\arraybackslash}m{1.3cm}|
    >{\columncolor{gray}\centering\arraybackslash}m{1.3cm}|
    >{\columncolor{cblue}\centering\arraybackslash}m{1.5cm}|
    >{\columncolor{gray}\centering\arraybackslash}m{1.8cm}|
    >{\columncolor{cblue}\centering\arraybackslash}m{1cm}|
    >{\columncolor{gray}\centering\arraybackslash}m{1cm}|
    >{\columncolor{cblue}\centering\arraybackslash}m{1cm}|
    >{\columncolor{gray}\centering\arraybackslash}m{0.8cm}|}

\caption{\textbf{Distance}-based methods \textbf{with public code} and their characteristics} \\
\hline
\rowcolor{white}
\textbf{Name} & \textbf{Subfamily} & \textbf{Dependency order} & \textbf{Continuous Time} & \textbf{Various Lengths} & \textbf{Missing Data} & \textbf{Multi Variable} & \textbf{Covariates} & \textbf{Main Data Type} & \multicolumn{3}{c|}{\textbf{Data Volume}} & \textbf{Ref.} \\
\cline{10-12}
\rowcolor{white}
 & & & & & & & & & \textbf{N} & \textbf{T} & \textbf{S} & \\
\hline
\endfirsthead

\hline
\rowcolor{white} 
\textbf{Name} & \textbf{Subfamily} & \textbf{Dependency order} & \textbf{Continuous Time} & \textbf{Various Lengths} & \textbf{Missing Data} & \textbf{Multivariate} & \textbf{Covariates} & \textbf{Data Type} & \multicolumn{3}{c|}{\textbf{Data Volume}} & \textbf{Ref.} \\
\cline{10-12}
\rowcolor{white}
 & & & & & & & & & \textbf{N} & \textbf{T} & \textbf{S} & \\
\hline
\endhead

\hline
\endfoot

Hamming Distance & Hamming & $0$ & $\times$ & $\times$ & $\times$ & $\times$ & $\times$ & Activities & $1200$ & $1400$ & $7$ & \cite{HammingMoreau}, \cite{HammingOriginal} \\ 
\hline

Dynamic Hamming Distance & Hamming & $1$ & $\times$ & $\times$ & $\times$ & $\times$ & $\times$ & Social & $8000$ & $150$ & $2$ & \cite{Distancereview}, \cite{DHD} \\ 
\hline

Fuzzy Temporal Hamming Distance & Hamming & $0$ & \checkmark & $\times$ & $\times$ & $\times$ & $\times$ & Activities & $1200$ & $1400$ & $7$ & \cite{HammingMoreau} \\ 
\hline

Damerau – Levenshtein Distance & OM & User & $\times$ & \checkmark & $\times$ & $\times$ & $\times$ & Care Trajectories & $1000$ & $15$ & $15$ & \cite{ModifiedNW}, \cite{Damerau1964} \\ 
\hline

Modified Needleman-Wunsch & OM & $\infty$ & $\times$ & \checkmark & $\times$ & $\times$ & $\times$ & Care trajectories & $1000$ & $15$ & $15$ & \cite{ModifiedNW} \\ 
\hline

Jaro Distance & OM & User & $\times$ & \checkmark & $\times$ & $\times$ & $\times$ & Care trajectories & $1000$ & $15$ & $15$ & \cite{ModifiedNW}, \cite{Jaro1989} \\ 
\hline

Clover & Selective Hamming & User & $\times$ & \checkmark & $\times$ & $\times$ & $\times$ & DNA & $10^{10}$ & $150$ & $4$ & \cite{Clover} \\ 
\hline

Randomized Discrete Sequence Clustering & Random search & $\infty$ & $\times$ & \checkmark & $\times$ & $\times$ & $\times$ & Clickstream & $5000$ & $7000$ & $3 \times 10^4$ & \cite{RandomizedAlgo} \\ 
\hline

Tribe-MCL & Selective OM & $\infty$ & $\times$ & \checkmark & $\times$ & $\times$ & $\times$ & Protein & $10^5$ & $400$ & $20$ & \cite{TribeMCL} \\ 
\hline

Contextual Edit Distance & OM & $\infty$ & $\times$ & \checkmark & $\times$ & $\times$ & $\times$ & Activities & $1200$ & $1400$ & $7$ & \cite{ContextualEdit} \\ 
\hline

GClust & LCS & $\infty$ & $\times$ & \checkmark & $\times$ & $\times$ & $\times$ & DNA & $10^5$ & $10^7$ & $4$ & \cite{Gclust} \\ 
\hline

Laplacian Eigenmaps & OM & $\infty$ & $\times$ & \checkmark & $\times$ & $\times$ & $\times$ & DNA & $30$ & $500$ & $4$ & \cite{LaplacianEigenmaps} \\ 
\hline

ProClust & OM (local) & $\infty$ & $\times$ & \checkmark & $\times$ & $\times$ & $\times$ & Protein & $6 \times 10^4$ & $350$ & $20$ & \cite{Proclust} \\ 
\hline

Cd-hit & Selective OM & $\infty$ & $\times$ & \checkmark & $\times$ & $\times$ & $\times$ & Protein & $6 \times 10^4$ & $500$ & $4$ & \cite{cdhit} \\ 
\hline

Slymfast & Random search & User & $\times$ & \checkmark & $\times$ & $\times$ & $\times$ & DNA & $2 \times 10^{7}$ & $100$ & $4$ & \cite{slymfast} \\ 
\hline

Jaro-Winkler Distance & OM & User & $\times$ & \checkmark & $\times$ & $\times$ & $\times$ & Care trajectories & $1000$ & $15$ & $15$ & \cite{ModifiedNW}, \cite{JaroWinkleroriginal} \\ 
\hline

Optimal Matching & OM & $\infty$ & $\times$ & \checkmark & $\times$ & $\times$ & $\times$ & Social & $300$ & $50$ & $35$ & \cite{Distancereview}, \cite{Abbott} \\ 
\hline

OMsloc & OM & $\infty$ & $\times$ & \checkmark & $\times$ & $\times$ & $\times$ & Social & $700$ & $100$ & $4$ & \cite{Distancereview}, \cite{OMsloc} \\ 
\hline

Longest Common Subsequence & LCS & $\infty$ & $\times$ & \checkmark & $\times$ & $\times$ & $\times$ & DNA & $7 \times 10^6$ & $75$ & $4$ & \cite{LLCS} \\ 
\hline

LinCLust & Selective OM & $\infty$ & $\times$ & \checkmark & $\times$ & $\times$ & $\times$ & Protein & $10^8$ & $400$ & $20$ & \cite{Linclust} \\ 
\hline

BlastClust & Selective OM & $\infty$ & $\times$ & \checkmark & $\times$ & $\times$ & $\times$ & DNA & $10^5$ & $10^7$ & $4$ &\cite{Gclust}, \cite{blastclust} \\ 
\hline

Extended Alphabet & OM & $\infty$ & $\times$ & \checkmark & $\times$ & \checkmark & $\times$ & Social & $1700$ & $26$ & $3 \times 5 \times 4 \times 2$ & \cite{Multichannelcomparison} \\ 
\hline

Multichannel Sequence Analysis & OM & $\infty$ & $\times$ & \checkmark & $\times$ & \checkmark & $\times$ & Social & $1800$ & $50$ & $7 \times 12 \times 10$ & \cite{Multichannelcomparison}, \cite{multichannel} \\ 
\hline

AliClu & OM & $\infty$ & \checkmark & \checkmark & $\times$ & $\times$ & $\times$ & Care trajectories & $426$ & $20$ & $8$ & \cite{aliclu} \\ 
\hline

GIMSA & OM & $\infty$ & $\times$ & \checkmark & $\times$ & \checkmark & $\times$ & Social & $1400$ & $50$ & $4 \times 5$ & \cite{GIMSA} \\ 
\hline

OMstran & OM & $\infty$ & $\times$ & \checkmark & \checkmark & $\times$ & $\times$ & Social & $2400$ & $20$ & $4$ & \cite{Distancereview}, \cite{OMstran} \\ 
\hline

\end{longtable}%

\endgroup

%% file: Tableaux/Tableaufeature.tex
\newpage
\newgeometry{top=2cm, bottom=2cm, left=0.5cm, right=0.5cm}

{ \scriptsize

\setlength{\extrarowheight}{1.15mm}

\begin{longtable}[c]{|
    >{\columncolor{gray}\centering\arraybackslash}m{1.8cm}|
    >{\columncolor{cred}\centering\arraybackslash}m{1.8cm}|
    >{\columncolor{gray}\centering\arraybackslash}m{1.8cm}|
    >{\columncolor{cred}\centering\arraybackslash}m{1.7cm}|
    >{\columncolor{gray}\centering\arraybackslash}m{1.3cm}|
    >{\columncolor{cred}\centering\arraybackslash}m{1.3cm}|
    >{\columncolor{gray}\centering\arraybackslash}m{1.3cm}|
    >{\columncolor{cred}\centering\arraybackslash}m{1.5cm}|
    >{\columncolor{gray}\centering\arraybackslash}m{1.8cm}|
    >{\columncolor{cred}\centering\arraybackslash}m{1cm}|
    >{\columncolor{gray}\centering\arraybackslash}m{1cm}|
    >{\columncolor{cred}\centering\arraybackslash}m{1cm}|
    >{\columncolor{gray}\centering\arraybackslash}m{0.8cm}|}

\caption{\textbf{Feature}-based methods \textbf{with public code} and their characteristics} \\
\hline
\rowcolor{white} 
\textbf{Name} & \textbf{Subfamily} & \textbf{Dependency order} & \textbf{Continuous Time} & \textbf{Various Lengths} & \textbf{Missing Data} & \textbf{Multi Variable} & \textbf{Covariates} & \textbf{Main Data Type} & \multicolumn{3}{c|}{\textbf{Data Volume}} & \textbf{Ref.} \\
\cline{10-12}
\rowcolor{white}
 & & & & & & & & & \textbf{N} & \textbf{T} & \textbf{S} & \\
\hline
\endfirsthead

\hline
\rowcolor{white} 
\textbf{Name} & \textbf{Subfamily} & \textbf{Dependency order} & \textbf{Continuous Time} & \textbf{Various Lengths} & \textbf{Missing Data} & \textbf{Multivariate} & \textbf{Covariates} & \textbf{Data Type} & \multicolumn{3}{c|}{\textbf{Data Volume}} & \textbf{Ref.} \\
\cline{10-12}
\rowcolor{white}
 & & & & & & & & & \textbf{N} & \textbf{T} & \textbf{S} & \\
\hline
\endhead

\hline
\endfoot

\hline
nTreeClus & Self-supervised (Random Forest)  & User & $\times$ & \checkmark & $\times$ & $\times$ & $\times$ & Social & $20 000$ & $90$ & $20$ & \cite{ntreeclus}\\ 
\hline

mBKM & Entropy & User & $\times$ & \checkmark & $\times$ & $\times$ & $\times$ & DNA & $4 \times 10^4$ & $100$ & $4$ & \cite{mBKM}\\ 
\hline

ISCT (Interpretable Sequence Clustering Tree) & kmers & $\infty$ & $\times$ & \checkmark & $\times$ & $\times$ & $\times$ & Clickstream & $5000$ & $7000$ & $3 \times 10^4$ & \cite{ISCT} \\ 
\hline

iDeLUCS & Chaos Game Representation & User & $\times$ & \checkmark & $\times$ & $\times$ & $\times$ & DNA & $3200$ & $5 \times 10^5 $ & $4$ & \cite{idelucs} \\ 
\hline

Sqn2Vec & Self-supervised learning (latent space)& User & $\times$ & \checkmark & $\times$ & $\times$ & $\times$ & Other & $5000$ & $3 \times 10^4$ & $7000$ & \cite{sqn2vec} \\ 
\hline

DeLUCS & Chaos Game Representation & User & $\times$ & \checkmark & $\times$ & $\times$ & $\times$ & DNA & $1200$ & $5 \times 10^5$& $4$ & \cite{DeLUCS}\\ 
\hline

STARS & Fourier & $\infty$ & $\times$ & \checkmark & $\times$ & $\times$ & $\times$ & DNA & $30$ & $3.4 \times 10^6$ & $4$ & \cite{STARS} \\
\hline

Envclust & Fourier & $\infty$ & $\times$ & \checkmark & $\times$ & $\times$ & $\times$ & Physiological & $80$& $1000$& $6$ & \cite{EnvClust} \\ 
\hline

MeShClust (v 3.0) & kmers & User & $\times$ & \checkmark & $\times$ & $\times$ & $\times$ & DNA & $10^4$ & $4 \times 10^6$& $4$ & \cite{Meshclustv3} \\ 
\hline

NMS (Number of matching subsequences) & SVR & $\infty$ & $\times$ & \checkmark & $\times$ & $\times$ & $\times$ & Social & $4300$& $15$& $8$ & \cite{NMSSVR} \\ 
\hline

MeShClust (v 1.0) & kmers & User & $\times$ & \checkmark & $\times$ & $\times$ & $\times$ & DNA & $100$& $10^4$& $4$ & \cite{MeshClust}\\ 
\hline

Categorical Functional Data Analysis (CFDA) & Functional data analysis & $\infty$ &  \checkmark & $\times$ & $\times$ & $\times$ & $\times$ & Care & $3000$&  $4$&  $10$ & \cite{CFDAapplied}, \cite{CFDApackage}\\ 
\hline

ctsfeatures & Correlation & User & $\times$ & \checkmark & $\times$ & $\times$ & $\times$ & Protein & $20$ & $7000$ & $4$ & \cite{ctsfeatures}, \cite{hardandsoft} \\ 
\hline

kmer & kmers & User & $\times$ & \checkmark & $\times$ & $\times$ & $\times$ & DNA & \textbf{NEI} & \textbf{NEI} & \textbf{NEI} &  \cite{kmerpackage}\\ 
\hline

CGR & Chaos Game Representation & $\infty$ &   $\times$ & \checkmark & $\times$ & $\times$ & $\times$ & Protein & \textbf{NEI} & \textbf{NEI} & \textbf{NEI} & \cite{CGR}\\ 
\hline

cluster Clickstreams & Markov & User &  $\times$ & \checkmark & \checkmark & $\times$ & $\times$ & Clickstream & $10^5$ & $50$ & $7$ & \cite{ClickstreamPackage} \\ 
\hline

Transition Matrix features & Markov & $1$ & $\times$ & \checkmark & $\times$ & \checkmark & $\times$ & Social & $120$ & $90$ & $(2 \times 2)$ & \cite{Transitionmatrixfeatures}\\ 
\hline

SVR (Subsequence Vector Representation) & SVR & $\infty$ &  \checkmark & \checkmark & $\times$ & $\times$ & $\times$ & Social & $4300$ & $15$ & $8$ & \cite{NMSSVR}\\ 
\hline

User Journey2Vector & Self-supervised learning (latent space) & User &  $\times$ & \checkmark & $\times$ & \checkmark & $\times$ & Clickstream & \textbf{NEI} & \textbf{NEI} & \textbf{NEI} & \cite{UserJourney}\\ 
\hline

\end{longtable}}

%% file: Tableaux/Tableaumodel.tex
\newpage
\newgeometry{top=2cm, bottom=2cm, left=0.5cm, right=0.5cm}

{ \scriptsize

\setlength{\extrarowheight}{1.15mm}
\begin{longtable}[c]{|
    >{\columncolor{gray}\centering\arraybackslash}m{1.8cm}|
    >{\columncolor{cgreen}\centering\arraybackslash}m{1.8cm}|
    >{\columncolor{gray}\centering\arraybackslash}m{1.8cm}|
    >{\columncolor{cgreen}\centering\arraybackslash}m{1.7cm}|
    >{\columncolor{gray}\centering\arraybackslash}m{1.3cm}|
    >{\columncolor{cgreen}\centering\arraybackslash}m{1.3cm}|
    >{\columncolor{gray}\centering\arraybackslash}m{1.3cm}|
    >{\columncolor{cgreen}\centering\arraybackslash}m{1.5cm}|
    >{\columncolor{gray}\centering\arraybackslash}m{1.8cm}|
    >{\columncolor{cgreen}\centering\arraybackslash}m{1cm}|
    >{\columncolor{gray}\centering\arraybackslash}m{1cm}|
    >{\columncolor{cgreen}\centering\arraybackslash}m{1cm}|
    >{\columncolor{gray}\centering\arraybackslash}m{0.8cm}|}

\caption{\textbf{Model}-based methods \textbf{with public code} and their characteristics} \\
\hline
\rowcolor{white} 
\textbf{Name} & \textbf{Subfamily} & \textbf{Dependency order} & \textbf{Continuous Time} & \textbf{Various Lengths} & \textbf{Missing Data} & \textbf{Multi Variable} & \textbf{Covariates} & \textbf{Main Data Type} & \multicolumn{3}{c|}{\textbf{Data Volume}} & \textbf{Ref.} \\
\cline{10-12}
\rowcolor{white}
 & & & & & & & & & \textbf{N} & \textbf{T} & \textbf{S} & \\
\hline
\endfirsthead

\hline
\rowcolor{white} 
\textbf{Name} & \textbf{Subfamily} & \textbf{Dependency order} & \textbf{Continuous Time} & \textbf{Various Lengths} & \textbf{Missing Data} & \textbf{Multivariate} & \textbf{Covariates} & \textbf{Data Type} & \multicolumn{3}{c|}{\textbf{Data Volume}} & \textbf{Ref.} \\
\cline{10-12}
\rowcolor{white}
 & & & & & & & & & \textbf{N} & \textbf{T} & \textbf{S} & \\
\hline
\endhead

\hline
\endfoot

\hline

Markov Chain Clustering & Markov & $1$ & $\times$ & \checkmark & $\times$ & $\times$ & $\times$ & Social & $10 000$ & $30$ & $6$ &\cite{bayesmarkov}, \cite{markovbayeslogit} \\ 
\hline

Dirichlet Multinomial Clustering & Markov & $\infty$  &  $\times$ & \checkmark & $\times$ & $\times$ & $\times$ &  Social & $10 000$ & $30$ & $6$ & \cite{bayesmarkov}\\ 
\hline

Clickclust & Markov & $1$ & $\times$ & \checkmark & $\times$ & $\times$ & $\times$ & Clickstream & $320$ & $360$ & $17$& \cite{biclustering},  \cite{Clickclust}\\ 
\hline

DBHC & HMM & $1$ & $\times$ & \checkmark & $\times$ & $\times$ & $\times$ & Social & $2000$ & $20$ & $8$ & \cite{DBHC}\\ 
\hline

MedSeq & Exponential & $0$ & $\times$ & $\times$ & $\times$ & $\times$ & \checkmark & Social & $700$ & $70$ & $6$ & \cite{Medseq} \\ 
\hline

LCA (Latent Class Analysis) & GLM & $0$ & $\times$ & $\times$ & \checkmark & \checkmark & $\times$ &  Social & $2290$ & $6$ & $32$ & \cite{LCAandSA}, \cite{polca} \\ 
\hline

Look-A-Liker (LAL) & RNN & $\infty$  & \checkmark & \checkmark & $\times$ & $\times$ & $\times$ & Clickstream & $10^6$&  $50$ & $5$ & \cite{lookaliker} \\ 
\hline

GBTM (Group-based Multi-Trajectory Modelling) & GLM & $0$ & $\times$ & $\times$ & $\times$ & \checkmark & \checkmark & Care trajectories & $10^6$ & $16$  & $2  \times 8$ & \cite{GBTMandSA}, \cite{GBTMoverview} \\ 
\hline

TESS (Temporal Event Sequence Summarization) & Combinatorial & $\infty$  & \checkmark & \checkmark & $\times$ & \checkmark & $\times$ & Other & $200$ &  $50 000$ &  $12$ & \cite{TESS}\\ 
\hline

 DMHP (Dirichlet Mixture of Hawkes Processes) & Hawkes & $\infty$  & \checkmark & \checkmark & \checkmark & $\times$ & $\times$ & Care trajectories & $2000$ & $50$ & $5$ & \cite{DMHP}\\ 
\hline

Mixture Hidden Markov Models & HMM  & $1$ & $\times$ & \checkmark & \checkmark & \checkmark & \checkmark & Social & $2000$ & $16$ & $3 \times 2 \times 2$ & \cite{seqHMM} \\ 
\hline

lamb (Latent factor Mixture model for Bayesian clustering) & GLM  & $\infty$& \checkmark & \checkmark & $\times$ & \checkmark & \checkmark & Physiological & $300$ & $1000$ & $10 \times 2$ & \cite{lamb} \\ 
\hline

Linear mixed-effect model & GLM & $\infty$ &  \checkmark & \checkmark & \checkmark & $\times$ & \checkmark & Social & $20 000$ & $4$ & $2$ & \cite{LinearMixedEffect}\\ 
\hline


Class-specific GLM & GLM & $\infty$ & \checkmark & \checkmark & \checkmark & $\times$ & \checkmark & Physiological & $750$ &  $126$ & $3$ & \cite{GLMclust} \\ 
\hline

\end{longtable}}

%% file: Tableaux/TableauDistancesAll.tex
\newpage
\newgeometry{top=2cm, bottom=2cm, left=0.5cm, right=0.5cm}

\begingroup
\scriptsize
\setlength{\extrarowheight}{1.15mm}

\begin{longtable}[c]{|
    >{\columncolor{gray}\centering\arraybackslash}m{1.8cm}|
    >{\columncolor{cblue}\centering\arraybackslash}m{1.8cm}|
    >{\columncolor{gray}\centering\arraybackslash}m{1.8cm}|
    >{\columncolor{cblue}\centering\arraybackslash}m{1.7cm}|
    >{\columncolor{gray}\centering\arraybackslash}m{1.3cm}|
    >{\columncolor{cblue}\centering\arraybackslash}m{1.3cm}|
    >{\columncolor{gray}\centering\arraybackslash}m{1.3cm}|
    >{\columncolor{cblue}\centering\arraybackslash}m{1.5cm}|
    >{\columncolor{gray}\centering\arraybackslash}m{1.8cm}|
    >{\columncolor{cblue}\centering\arraybackslash}m{1cm}|
    >{\columncolor{gray}\centering\arraybackslash}m{1cm}|
    >{\columncolor{cblue}\centering\arraybackslash}m{1cm}|
    >{\columncolor{gray}\centering\arraybackslash}m{0.8cm}|}

\caption{All found \textbf{Distance}-based methods and their characteristics} \\
\hline
\rowcolor{white}
\textbf{Name} & \textbf{Subfamily} & \textbf{Dependency order} & \textbf{Continuous Time} & \textbf{Various Lengths} & \textbf{Missing Data} & \textbf{Multi Variable} & \textbf{Covariates} & \textbf{Main Data Type} & \multicolumn{3}{c|}{\textbf{Data Volume}} & \textbf{Ref.} \\
\cline{10-12}
\rowcolor{white}
 & & & & & & & & & \textbf{N} & \textbf{T} & \textbf{S} & \\
\hline
\endfirsthead

\hline
\rowcolor{white} 
\textbf{Name} & \textbf{Subfamily} & \textbf{Dependency order} & \textbf{Continuous Time} & \textbf{Various Lengths} & \textbf{Missing Data} & \textbf{Multi Variable} & \textbf{Covariates} & \textbf{Main Data Type} & \multicolumn{3}{c|}{\textbf{Data Volume}} & \textbf{Ref.} \\
\cline{10-12}
\rowcolor{white}
 & & & & & & & & & \textbf{N} & \textbf{T} & \textbf{S} & \\
\hline
\endhead

\hline
\endfoot

$d_{\text{Corr}}$ : Correlation distance & Correlation & $1$ & $\times$ & $\times$  & $\times$ & $\times$ & $\times$ & Clickstream & $10^4$ & $30$ & $6$ & \cite{garcia2015framework} \\ 
\hline

DHD (Dynamic Hamming distance) & Hamming & $1$ & $\times$ & $\times$ & $\times$ & $\times$ & $\times$ & Social & $8000$ & $150$ & $2$ & \cite{Distancereview}, \cite{DHD} \\ 
\hline

Hamming distance & Hamming & $0$ & $\times$ & $\times$ & $\times$ & $\times$ & $\times$ & Activities & $1200$ & $1400$ & $7$ & \cite{HammingMoreau}, \cite{hamming1950error} \\ 
\hline

Multiple Relational Self Organizing Map (MR - SOM) & OM & $\infty$ & $\times$ & \checkmark & $\times$ & $\times$ & $\times$ & Social & $2 \times 10^4$ & $100$ & $9$ & \cite{SOM}\\ 
\hline

Cd-hit & Selective OM & $\infty$ & $\times$ & \checkmark & $\times$ & $\times$ & $\times$ & Protein & $6 \times 10^6$ & $500$ & $4$ & \cite{cdhit}\\ 
\hline

ProClust & OM (local) & $\infty$ & $\times$ & \checkmark & $\times$ & $\times$ & $\times$ & Protein & $6 \times 10^4$ & $350$ & $20$ &\cite{Proclust} \\ 
\hline

Tribe-MCL (Markov cluster) & BLAST & $\infty$ & $\times$ & \checkmark & $\times$ & $\times$ & $\times$ & Protein & $10^5$ & $400$ & $20$ & \cite{TribeMCL} \\ 
\hline 

LinCLust & Selective OM & $\infty$ & $\times$ & \checkmark & $\times$ & $\times$ & $\times$ & Protein & $10^8$ & $400$ & $4$ & \cite{Linclust} \\ 
\hline

SCMDNS, SCLARANS & OM & $\infty$ & $\times$ & \checkmark & $\times$ & $\times$ & $\times$ & Protein & $1400$ & $500$ & $20$ & \cite{SCLARANS} \\ 
\hline

Optimal Matching-based fuzzy-possibilistic c-medoids  & OM & $\infty$ & $\times$ & \checkmark & $\times$ & $\times$ & $\times$ & Protein & \textbf{NEI} & \textbf{NEI} & \textbf{NEI} & \cite{cmedoids}\\ 
\hline

POPC (Pattern-Oriented Partial Clustering) & Jaccard & $\infty$ & $\times$ & \checkmark & $\times$ & $\times$ & $\times$ & Other & $10$ & $5$ & $10$ & \cite{popc} \\ 
\hline

BAG & OM & $\infty$ & $\times$ & \checkmark & $\times$ & $\times$ & $\times$ & Other & \textbf{NEI} & \textbf{NEI} & \textbf{NEI} & \cite{BAG} \\ 
\hline

Optimal Matching & OM & $\infty$ & $\times$ & \checkmark & $\times$ & $\times$ & $\times$ & Social & $300$ & $50$ & $35$ & \cite{Distancereview}, \cite{Abbott}, \cite{Distancereview}\\ 
\hline

Laplacian Eigenmaps & OM & $\infty$ & $\times$ & \checkmark & $\times$ & $\times$ & $\times$ & DNA & $30$ & $500$ & $4$ & \cite{LaplacianEigenmaps}\\ 
\hline

BlastClust & BLAST & $\infty$ & $\times$ & \checkmark & $\times$ & $\times$ & $\times$ & DNA & $10^5$& $10^7$ & $4$ & \cite{blastclust} \\ 
\hline

LLCS (Length of the Longest Common Subsequence) & LCS & $\infty$ & $\times$ & \checkmark & $\times$ & $\times$ & $\times$ & DNA & $7 \times 10^6$ & $75$ & $4$ & \cite{LLCS}\\ 
\hline

k-means for event sequences & OM & $\infty$ & $\times$ & \checkmark & $\times$ & $\times$ & $\times$ & Other & $2400$ & $78$ & $285$ & \cite{SequentialClust} \\ 
\hline

OMsloc & OM & $\infty$ & $\times$ & \checkmark & $\times$ & $\times$ & $\times$ & Social & $700$ & $100$ & $4$ & \cite{OMsloc} \\ 
\hline

Kullback-Leibler divergence for Markovian sequences & KL & User & $\times$ & \checkmark & $\times$ & $\times$ & $\times$ & Other & $16$ & $5000$ & $30$ & \cite{Agnostic}\\ 
\hline

Slymfast & Random search & User & $\times$ & \checkmark & $\times$ & $\times$ & $\times$ & DNA & $2 \times 10^7$ & $100$ & $4$ & \cite{slymfast}\\ 
\hline

Clover & Selective Hamming & User & $\times$ & \checkmark & $\times$ & $\times$ & $\times$ & DNA & $10^{10}$ & $150$ & $4$ & \cite{Clover}\\ 
\hline

Entropy-based Kullback-Leibler divergence & KL & User & $\times$ & \checkmark & $\times$ & $\times$ & $\times$ & DNA & $30$ & $8\times 10^4$ & $4$ & \cite{KLentropy}\\ 
\hline

Jaro Distance & OM & User & $\times$ & \checkmark & $\times$ & $\times$ & $\times$ & Care trajectories & $1000$ & $15$ & $15$ & \cite{ModifiedNW}, \cite{Jaro1989}\\ 
\hline

Damerau– Levenshtein Distance & OM & User & $\times$ & \checkmark & $\times$ & $\times$ & $\times$ & Care trajectories & $1000$ & $15$ & $15$ & \cite{ModifiedNW}, \cite{Damerau1964}\\ 
\hline

Jaro-Winkler Distance & OM & User & $\times$ & \checkmark & $\times$ & $\times$ & $\times$ & Care trajectories & $1000$ & $15$ & $15$ & \cite{ModifiedNW}, \cite{JaroWinkleroriginal}\\ 
\hline

Rank Distance & Rank & $\infty$ & $\times$ & \checkmark & $\times$ & $\times$ & $\times$ & DNA & $22$ & $2 \times 10^4$ & $4$ & \cite{Rank} \\ 
\hline

k-means for DNA sequences & OM & $\infty$ & $\times$ & \checkmark & $\times$ & $\times$ & $\times$ & DNA & \textbf{NEI} & \textbf{NEI} & \textbf{NEI} & \cite{improved} \\  
\hline

GClust & LCS & $\infty$ & $\times$ & \checkmark & $\times$ & $\times$ & $\times$ & DNA & $10^5$& $10^7$ & $4$ & \cite{Gclust} \\ 
\hline

Scalable Jaccard-type Distance & Jaccard & $\infty$ & $\times$ & \checkmark & $\times$ & $\times$ & $\times$ & DNA & $1500$ & $60$ & $4$ & \cite{scalable2005} \\ 
\hline

LAHDC (Landmark-based Active Hierarchical Divisive Clustering) & Selective OM & $\infty$ & $\times$ & \checkmark & $\times$ & $\times$ & $\times$ & DNA & $7 \times 10^6$ & $500$ & $4$ & \cite{LAHDC} \\ 
\hline

Markov Fuzzy integral similarity & Markov & $1$ & $\times$ & \checkmark & $\times$ & $\times$ & $\times$ & DNA & 30 & $8 \times 10^6$ & $4$ &  \cite{Fuzzyintegral} \\ 
\hline

SimMM (Similarity of Markov Models) & KL & $1$ & $\times$ & \checkmark & $\times$ & $\times$ & $\times$ & DNA & $20$ & $14000$ & $4$ & \cite{SimMM} \\ 
\hline

PST (probabilistic suffix tree)-based Jaccard distance  & Jaccard & $\infty$ & $\times$ & \checkmark & $\times$ & $\times$ & $\times$ & Activities & \textbf{NEI} & \textbf{NEI} & \textbf{NEI} &  \cite{Patternbased} \\ 
\hline

Contextual Edit Distance & OM & $\infty$ & $\times$ & \checkmark & $\times$ & $\times$ & $\times$ & Activities & $1200$ & $1400$ & $7$ & \cite{ContextualEdit} \\ 
\hline

KCET (Kmeans Clustering  through Evolutionary Templates) & OM & $\infty$ & $\times$ & \checkmark & $\times$ & $\times$ & $\times$ & Care trajectories & $3500$ & $10$ & $6$ & \cite{Towards} \\  
\hline

Modified Needleman-Wunsch & OM & $\infty$ & $\times$ & \checkmark & $\times$ & $\times$ & $\times$ & Care trajectories & $1000$ & $15$ & $15$ & \cite{ModifiedNW} \\ 
\hline

Jaccard distance & Jaccard & $\infty$ & $\times$ & \checkmark & $\times$ & $\times$ & $\times$ & DNA & $1500$ & $60$ & $4$ & \cite{jaccardhierarchical} \\ 
\hline

Fuzzy Temporal Hamming (FTH) & Hamming & $0$ & \checkmark & $\times$ & $\times$ & $\times$ & $\times$ & Activities & $1200$ & $1400$ & $7$ & \cite{HammingMoreau}\\  
\hline

RFSC (Random Forest for Discrete Sequence Clustering) & Self-supervised learning (Random Forest) & $\infty$ & $\times$ & \checkmark & $\times$ & $\times$ & $\times$ & Clickstream & $5000$ & $7000$ & $30000$ & \cite{RFSC} \\ 
\hline

Mixed Panel New distance & Hamming & $0$ & $\times$ & $\times$ & \checkmark & $\times$ & $\times$ & Other & $25$ & $12$ & $2 \times 2 \times 2  \times 2 \times 2 \times 2$ & \cite{Panelnewdistance} \\   
\hline

RSC (Randomized Discrete Sequence Clustering) & Random search & $\infty$ & $\times$ & \checkmark & $\times$ & $\times$ & $\times$ & Clickstream & $5000$ & $7000$ & $3 \times 10^4$ & \cite{RandomizedAlgo} \\  
\hline

GIMSA (Global Interdependence Approach to Multidimensional Sequence Analysis) & OM & $\infty$ & $\times$ & \checkmark & $\times$ & \checkmark & $\times$ & Social & $1400$ & $50$ & $4\times5$ & \cite{GIMSA} \\ 
\hline

Gower Distance & Hamming & $0$ & $\times$ & $\times$ & \checkmark & \checkmark & $\times$ & Other & $25$ & $12$ & $2\times 2 \times 2\times 2\times 2\times 2$ &  \cite{panelgower} \\ 
\hline

AliClu & OM & $\infty$ & \checkmark & \checkmark & $\times$ & $\times$ & $\times$ & Care trajectories & $426$ & $20$ & $8$ & \cite{aliclu} \\  
\hline

DTW & Dynamic Time Warping & $\infty$ & \checkmark & \checkmark & $\times$ & $\times$ & $\times$ & Care trajectories & $7500$ & $8$ & $11$ & \cite{DTWcare} \\ 
\hline

CombT (merged combined domain types) & OM & $\infty$ & $\times$ & \checkmark & $\times$ & \checkmark & $\times$ & Social & $2000$ & $50$ & $9 \times 10 \times 5$ &  \cite{strategies} \\ 
\hline

Multichannel Optimal Matching-based Fuzzy Clustering & OM & $\infty$ & $\times$ & \checkmark & $\times$ & \checkmark & $\times$ & Social & $6800$ & $200$ & $7\times 4$ & \cite{tools} \\ 
\hline

ACS (all common subsequence) similarity & LCS & $\infty$ & $\times$ & \checkmark & $\times$ & \checkmark & $\times$ & Care trajectories & $1000$ & $25$ & $2^{30}$ & \cite{ACS}\\  
\hline

Kullback-Leibler Divergence on HMM & KL & $\infty$ & $\times$ & \checkmark & $\times$ & \checkmark & $\times$ & Clinical & $1300$ & $20$ & $2 \times 2 \times 2 \times 4 \times 3$ & \cite{HMMClust} \\   
\hline

OMstran & OM & $\infty$ & $\times$ & \checkmark & \checkmark & $\times$ & $\times$ & Social & $2400$ & $20$ & $4$ & \cite{Distancereview}, \cite{OMstran}\\ 
\hline

MSA (Multichannel Sequence Analysis) & OM & $\infty$ & $\times$ & \checkmark & $\times$ & \checkmark & $\times$ & Social & $1800$ & $50$ & $7 \times 12 \times 10$ & \cite{multichannel}, \cite{Multichannelcomparison} \\  
\hline

EA (Extended Alphabet) & OM & $\infty$ & $\times$ & \checkmark & $\times$ & \checkmark & $\times$ & Social & $1700$ & $26$ & $3\times 5 \times 4 \times 2$ & \cite{Multichannelcomparison} \\  
\hline

Optimal Matching-based interpretable Divisive Clustering & OM  & $\infty$ & \checkmark & \checkmark & $\times$ & $\times$ & $\times$ & Social & $600$ & $200$ & $16$ & \cite{DivisiveSocial} \\  
\hline

Regression tree-based discrepancy analysis & LCS & $\infty$ & $\times$ & \checkmark & $\times$ & $\times$ & \checkmark & Care trajectories & $5600$ & $50$ & $6$ & \cite{lemeur} \\ 
\hline

Multichannel Optimal Matching with Missing Values & OM & $\infty$ & $\times$ & \checkmark & $\checkmark$ & \checkmark & $\times$ & Social & $1000$ & $60$ & $10 \times 10 \times 2 \times 2 \times 2 $ &  \cite{dataquality} \\ 
\hline

Hierastiseq & Dynamic Time Warping & $\infty$ & \checkmark & \checkmark & \checkmark & $\times$ & $\times$ & Care trajectories & $3300$ & $90$ & $13$ & \cite{hierastiseq} \\  
\hline
\end{longtable}%

\endgroup

%% file: Tableaux/TableauFeatureAll.tex
\newpage
\newgeometry{top=2cm, bottom=2cm, left=0.5cm, right=0.5cm}

{ \scriptsize

\setlength{\extrarowheight}{1.15mm}

\begin{longtable}[c]{|
    >{\columncolor{gray}\centering\arraybackslash}m{1.8cm}|
    >{\columncolor{cred}\centering\arraybackslash}m{1.8cm}|
    >{\columncolor{gray}\centering\arraybackslash}m{1.8cm}|
    >{\columncolor{cred}\centering\arraybackslash}m{1.7cm}|
    >{\columncolor{gray}\centering\arraybackslash}m{1.3cm}|
    >{\columncolor{cred}\centering\arraybackslash}m{1.3cm}|
    >{\columncolor{gray}\centering\arraybackslash}m{1.3cm}|
    >{\columncolor{cred}\centering\arraybackslash}m{1.5cm}|
    >{\columncolor{gray}\centering\arraybackslash}m{1.8cm}|
    >{\columncolor{cred}\centering\arraybackslash}m{1cm}|
    >{\columncolor{gray}\centering\arraybackslash}m{1cm}|
    >{\columncolor{cred}\centering\arraybackslash}m{1cm}|
    >{\columncolor{gray}\centering\arraybackslash}m{0.8cm}|}

\caption{All found \textbf{Feature}-based methods and their characteristics} \\
\hline
\rowcolor{white} 
\textbf{Name} & \textbf{Subfamily} & \textbf{Dependency order} & \textbf{Continuous Time} & \textbf{Various Lengths} & \textbf{Missing Data} & \textbf{Multi Variable} & \textbf{Covariates} & \textbf{Main Data Type} & \multicolumn{3}{c|}{\textbf{Data Volume}} & \textbf{Ref.} \\
\cline{10-12}
\rowcolor{white}
 & & & & & & & & & \textbf{N} & \textbf{T} & \textbf{S} & \\
\hline
\endfirsthead

\hline
\rowcolor{white} 
\textbf{Name} & \textbf{Subfamily} & \textbf{Dependency order} & \textbf{Continuous Time} & \textbf{Various Lengths} & \textbf{Missing Data} & \textbf{Multi Variable} & \textbf{Covariates} & \textbf{Data Type} & \multicolumn{3}{c|}{\textbf{Data Volume}} & \textbf{Ref.} \\
\cline{10-12}
\rowcolor{white}
 & & & & & & & & & \textbf{N} & \textbf{T} & \textbf{S} & \\
\hline
\endhead

\hline
\endfoot

Entropy-based online hierarchical tree & Frequencies & $0$ & $\times$ & $\times$ & $\times$ & $\times$ & $\times$ & Clickstream  & $20$ & $10^5$ & $70$  & \cite{HCnominalDatastreams} \\ 
\hline

Persistence landscapes of Walsh Fourier transform & Fourier transform & $\infty$ & $\times$ & $\times$ & $\times$ & $\times$ & $\times$ & Activities & $2.5 \times 10^5$ & $1400$ & $3$ & \cite{TDACTS}\\  
\hline

KM-NVLT (Kmeans with Normalized Variable-length Tuples) & kmers & User & $\times$ & \checkmark & $\times$ & $\times$ & $\times$ & Speech & $50$ & $3800$ & $18$ & \cite{tuplelength}\\  
\hline

nTreeClus & Self-supervised learning (Random Forest) & User & $\times$ & \checkmark & $\times$ & $\times$ & $\times$ & Social & $2 \times 10^4$ & $90$ & $20$ & \cite{ntreeclus}\\ 
\hline

mBKM & Entropy & User & $\times$ & \checkmark & $\times$ & $\times$ & $\times$ & DNA & $4 \times 10^4$ & $100$ & $4$ & \cite{mBKM}\\  
\hline

ISCT (Interpretable Sequence Clustering Tree) & kmers & $\infty$ & $\times$ & \checkmark & $\times$ & $\times$ & $\times$ & Clickstream & $5000$ & $7000$ & $28000$ & \cite{ISCT}\\ 
\hline

iDeLUCS & Chaos Game Representation & User & $\times$ & \checkmark & $\times$ & $\times$ & $\times$ & DNA & $3200$ & $5 \times 10^5$ & $4$ & \cite{idelucs}\\ 
\hline

Hamming distance on kmers vectors, with SVM-based outlier detection & kmers & User & $\times$ & \checkmark & $\times$ & $\times$ & $\times$ & DNA & $100$ & $500$ & $4$ &  \cite{DeNovo}\\  
\hline

Four approaches (local/global, frequent/long patterns) for kmers reduction & kmers & User & $\times$ & \checkmark & $\times$ & $\times$ & $\times$ & Other & $44000$ & $400$ & $20$ & \cite{Ascalablealg}\\  
\hline

Out-of-place distance on kmers & kmers & User & $\times$ & \checkmark & $\times$ & $\times$ & $\times$ & DNA & $50$ & $18000$ & $4$ & \cite{Outofplace}\\  
\hline

Sqn2Vec & Self-supervised learning (latent space) & User & $\times$ & \checkmark & $\times$ & $\times$ & $\times$ & Other & $5000$ & $28000$ & $7000$ & \cite{sqn2vec}\\  
\hline

STARS & Fourier transform & $\infty$ & $\times$ & \checkmark & $\times$ & $\times$ & $\times$ & DNA & $30$ & $3.4 \times 10^6$ & $4$ & \cite{STARS} \\
\hline

Sequen-C & kmers & User & $\times$ & \checkmark & $\times$ & $\times$ & $\times$ & Care trajectories & $22000$ & $180$ & $11$ &  \cite{Sequen-C}\\
\hline

Entropy-based kmers reduction & kmers & User & $\times$ & \checkmark & $\times$ & $\times$ & $\times$ & DNA & $800$ & $3700$ & $5$ & \cite{Twostagepruning}\\  
\hline

MeShClust (v 1.0) & kmers & User  & $\times$ & \checkmark & $\times$ & $\times$ & $\times$ & DNA & $100$ & $13000$ & $4$ &\cite{MeshClust}\\ 
\hline

kmer & kmers & User & $\times$ & \checkmark & $\times$ & $\times$ & $\times$ & DNA & \textbf{NEI} & \textbf{NEI} & \textbf{NEI} & \cite{kmerpackage}\\  
\hline

ctsfeatures & Correlation & User & $\times$ & \checkmark & $\times$ & $\times$ & $\times$ & Protein & $20$ & $7000$ & $4$ & \cite{ctsfeatures}, \cite{hardandsoft}\\  
\hline

Categorical Functional Data Analysis (CFDA) & Functional data analysis & $\infty$ & \checkmark & $\times$ & $\times$ & $\times$ & $\times$ & Care trajectories & $3000$ & $4$ & $10$ & \cite{CFDAapplied}, \cite{CFDApackage}\\  
\hline

NMS (Number of matching subsequences) & SVR & $\infty$ & $\times$ & \checkmark & $\times$ & $\times$ & $\times$ & Social & $4300$ & $15$ & $8$ & \cite{NMSSVR}, \cite{Distancereview}\\ 
\hline

DOMM (Dynamic Order Markov Model)-based vectorization & Markov & $\infty$ & $\times$ & \checkmark & $\times$ & $\times$ & $\times$ & Speech & $2000$ & $1700$ & $20$ & \cite{DOMM}\\  
\hline

RKMS (Robust K -means for sequences) & kmers & $\infty$ & $\times$ & \checkmark & $\times$ & $\times$ & $\times$ & Speech & $800$ & $3500$ & $5$ & \cite{RKMS}\\ 
\hline

MeShClust (v 3.0) & kmers & User & $\times$ & \checkmark & $\times$ & $\times$ & $\times$ & DNA & $10^4$ & $4 \times 10^6$ & $4$ & \cite{Meshclustv3}\\ 
\hline

DHCS (Divisive Hierarchical Clustering algorithm for categorical sequences) & Markov & $\infty$ & $\times$ & \checkmark & $\times$ & $\times$ & $\times$ & Speech & $1200$ & $1700$ & $20$ & \cite{DHCS}\\ 
\hline

DFT (Discrete Fourier transform)  & Fourier transform & $\infty$ & $\times$ & \checkmark & $\times$ & $\times$ & $\times$ & DNA & $80$ & $7000$ & $4$ & \cite{FourierTransform}\\ 
\hline

Wavelet-based Feature Vector (WFV) & Wavelets & $\infty$ & $\times$ & \checkmark & $\times$ & $\times$ & $\times$ & DNA & $28000$ & $1400$ & $4$ &  \cite{Waveletbased}\\
\hline

Moments-based feature vectors & Moments & $\infty$ & $\times$ & \checkmark & $\times$ & $\times$ & $\times$ & DNA & $4$ & $435$ & $4$ & \cite{FeatureMoments}\\  
\hline

Randomly selected N-grams & kmers & $\infty$ & $\times$ & \checkmark & $\times$ & $\times$ & $\times$ & DNA & \textbf{NEI} & \textbf{NEI} & \textbf{NEI} & \cite{NgramsDNA}\\  
\hline

HFD (Higuchi Fractal Dimension) & Fractal dim & $\infty$ & $\times$ & \checkmark & $\times$ & $\times$ & $\times$ & DNA & \textbf{NEI} & \textbf{NEI} & \textbf{NEI} & \cite{FractalDimension}\\  
\hline

DNA-MC (DNA Mapping and Clustering) & Fourier transform & $\infty$ & $\times$ & \checkmark & $\times$ & $\times$ & $\times$ & DNA & \textbf{NEI} & \textbf{NEI} & \textbf{NEI} &  \cite{DNAMC}\\  
\hline

WNN/PS(Wavelet Neural Network/Power Spectrum) & Fourier transform & $\infty$ & $\times$ & \checkmark & $\times$ & $\times$ & $\times$ & DNA & \textbf{NEI} & \textbf{NEI} & \textbf{NEI} &  \cite{WaveletNN}\\   
\hline

RFT (Ramanujan-Fourier Transform) & Fourier transform & $\infty$ & $\times$ & \checkmark & $\times$ & $\times$ & $\times$ & DNA & $500$ & $650$ & $4$ & \cite{RamanujanFourier}\\ 
\hline

Latent Dirichlet model-based feature vectors & Latent Dirichlet Model & $\infty$ & $\times$ & \checkmark & $\times$ & $\times$ & $\times$ & Care trajectories & $300$ & $40$ & $300$ & \cite{PatientTraces} \\  
\hline

CGR & Chaos Game Representation & $\infty$ & $\times$ & \checkmark & $\times$ & $\times$ & $\times$ & Protein & \textbf{NEI} & \textbf{NEI} & \textbf{NEI} & \cite{CGR}\\  
\hline

MKCC (Multi-view Clustering for Categorical seqences) & HMM & $1$ & $\times$ & \checkmark & $\times$ & $\times$ & $\times$ & Speech & \textbf{NEI} & \textbf{NEI} & \textbf{NEI} & \cite{Multiview}\\  
\hline

DeLUCS & Chaos Game Representation & User & $\times$ & \checkmark & $\times$ & $\times$ & $\times$ & DNA &  $1200$ & $4.7 \times 10^5$ & $4$ & \cite{DeLUCS}\\ 
\hline

Envclust & Fourier transform & $\infty$ & $\times$ & \checkmark & $\times$ & $\times$ & $\times$ & Clinical & $80$ & $1000$ & $6$ & \cite{EnvClust}\\   
\hline

FHC-NDS & Frequencies & $0$ & $\times$ & \checkmark & $\times$ & $\times$ & $\times$ & Clickstream & $30000$ & $10$ & $24$ & \cite{FHC-NDS}\\
\hline

CPF (Category-Position-Frequency) & Entropy & $1$ & $\times$ & \checkmark & $\times$ & $\times$ & $\times$ & DNA & $10000$ & $1500$ & $4$ & \cite{Animproved}\\  
\hline

Query-based kmers reduction & kmers & User & $\times$ & \checkmark & $\times$ & $\times$ & $\times$ & Activities & $460$& $1440$ & $7$ & \cite{Arewe}\\   
\hline

Multichannel transition Matrix features & Markov & $1$ & $\times$ & \checkmark & $\times$ & \checkmark & $\times$ & Social & $120$ & $90$ & $2 \times 2$ & \cite{Transitionmatrixfeatures}\\ 
\hline

User Journey2Vector & Self-supervised learning (Transformer) & User & $\times$ & \checkmark & $\times$ & \checkmark & $\times$ & Clickstream & \textbf{NEI} & \textbf{NEI} & \textbf{NEI} & \cite{UserJourney}\\ 
\hline

Transition frequency-based feature vectors & Markov & $1$ & $\times$ & \checkmark & $\times$ & \checkmark & $\times$ & Care trajectories & $160$ & $70$ & $2$ & \cite{MiningClinicam}\\  
\hline

Transition matrix-based feature vectors & Markov & $1$ & $\times$ & \checkmark & $\times$ & \checkmark & $\times$ & Care trajectories & $5000$ & $50$ & $10$ & \cite{Typical}\\ 
\hline

cluster Clickstreams & Markov & User & $\times$ & \checkmark & \checkmark & $\times$ & $\times$ & Clickstream & $10^5$ & $50$ & $7$ & \cite{ClickstreamPackage}\\  
\hline

SVR (Subsequence Vector Representation) & SVR & $\infty$ & \checkmark & \checkmark & $\times$ & $\times$ & $\times$ & Social & $4300$ & $15$ & $8$ & \cite{Distancereview}, \cite{NMSSVR}\\  
\hline

\end{longtable}}

%% file: Tableaux/Tableaumodelall.tex
\newpage
\newgeometry{top=2cm, bottom=2cm, left=0.5cm, right=0.5cm}

\begingroup
\scriptsize
\setlength{\extrarowheight}{1.15mm}

\begin{longtable}[c]{|
    >{\columncolor{gray}\centering\arraybackslash}m{1.8cm}|
    >{\columncolor{cgreen}\centering\arraybackslash}m{1.8cm}|
    >{\columncolor{gray}\centering\arraybackslash}m{1.8cm}|
    >{\columncolor{cgreen}\centering\arraybackslash}m{1.7cm}|
    >{\columncolor{gray}\centering\arraybackslash}m{1.3cm}|
    >{\columncolor{cgreen}\centering\arraybackslash}m{1.3cm}|
    >{\columncolor{gray}\centering\arraybackslash}m{1.3cm}|
    >{\columncolor{cgreen}\centering\arraybackslash}m{1.5cm}|
    >{\columncolor{gray}\centering\arraybackslash}m{1.8cm}|
    >{\columncolor{cgreen}\centering\arraybackslash}m{1cm}|
    >{\columncolor{gray}\centering\arraybackslash}m{1cm}|
    >{\columncolor{cgreen}\centering\arraybackslash}m{1cm}|
    >{\columncolor{gray}\centering\arraybackslash}m{0.8cm}|}

\caption{All found \textbf{Model}-based methods and their characteristics} \\
\hline
\rowcolor{white}
\textbf{Name} & \textbf{Subfamily} & \textbf{Dependency order} & \textbf{Continuous Time} & \textbf{Various Lengths} & \textbf{Missing Data} & \textbf{Multi Variable} & \textbf{Covariates} & \textbf{Main Data Type} & \multicolumn{3}{c|}{\textbf{Data Volume}} & \textbf{Ref.} \\
\cline{10-12}
\rowcolor{white}
 & & & & & & & & & \textbf{N} & \textbf{T} & \textbf{S} & \\
\hline
\endfirsthead

\hline
\rowcolor{white} 
\textbf{Name} & \textbf{Subfamily} & \textbf{Dependency order} & \textbf{Continuous Time} & \textbf{Various Lengths} & \textbf{Missing Data} & \textbf{Multi Variable} & \textbf{Covariates} & \textbf{Main Data Type} & \multicolumn{3}{c|}{\textbf{Data Volume}} & \textbf{Ref.} \\
\cline{10-12}
\rowcolor{white}
 & & & & & & & & & \textbf{N} & \textbf{T} & \textbf{S} & \\
\hline
\endhead

\hline
\endfoot

Panel HMM-based KL distance & HMM & $1$ & $\times$ & \checkmark & $\times$ & $\times$ & $\times$ & Clickstream & 2000 & 91 & 10 & \cite{Miningcategorical}\\ 
\hline

Divide and Combine-based Mixture Markov models & Markov & $1$ & $\times$ & \checkmark & $\times$ & $\times$ & $\times$ & Activities & $2,5 \times 10^5$ & $1400$ & $5$ & \cite{Divideandcombine}\\  
\hline

Nonhomogeneous first order Markov chain model & Markov & 0 or 1 & $\times$ & \checkmark & $\times$ & $\times$ & $\times$ & DNA & $350$ & $70$ & $4$ & \cite{Nonhomogeneous}\\ 
\hline

MedSeq & Exponential & $0$ & $\times$ & $\times$ & $\times$ & $\times$ & \checkmark & Social & $700$ & $70$ & $6$ & \cite{Medseq}\\  
\hline

DBHC & HMM & $1$ & $\times$ & \checkmark & $\times$ & $\times$ & $\times$ & Social & $2000$ & $20$ & $8$ & \cite{DBHC}\\  
\hline

Clickclust & Markov & $1$ & $\times$ & \checkmark & $\times$ & $\times$ & $\times$ & Clickstream & $320$ & $360$ & $17$ & \cite{Clickclust}\\  
\hline

Arbitrary order Markov-model-based greedy join and split algorithm & Markov & User & $\times$ & \checkmark & $\times$ & $\times$ & $\times$ & DNA & $2 \times 10^5$ & $650$ & $4$ & \cite{BayesianClustering}\\  
\hline

Dirichlet Multinomial Clustering & Markov & $\infty$ & $\times$ & \checkmark & $\times$ & $\times$ & $\times$ & Social & $10000$ & $30$ & $6$ & \cite{bayesmarkov}\\  
\hline

Markov Chain Clustering & Markov & $1$ & $\times$ & \checkmark & $\times$ & $\times$ & $\times$ & Social & $10000$ & $30$ & $6$ & \cite{bayesmarkov}\\  
\hline

GBTM (Group-based Multi-Trajctory Modelling) & GLM & $0$ & $\times$ & $\times$ & $\times$ & \checkmark & \checkmark & Care trajectories & $10^6$ & $16$ & $2\times8$ & \cite{GBTMoverview}, \cite{GBTMandSA}\\  
\hline

Look-A-Liker (LAL) & RNN & $\infty$ & \checkmark & \checkmark & $\times$ & $\times$ & $\times$ & Clickstream & $10^6$ & $50$ & $5$ &\cite{lookaliker}\\ 
\hline

Semi-Markov Model-based clustering & Markov & $1$ & \checkmark & \checkmark & $\times$ & $\times$ & $\times$ & Care trajectories & $10000$ & $100$ & $5$ & \cite{Impactofestimation}\\  
\hline

First-order Markov model with secondary components & Markov & $1$ & $\times$ & \checkmark & \checkmark & $\times$ & $\times$ & Clickstream & $300$ & $400$ & $17$ &\cite{Secondarycomponents}\\  
\hline

Combinatorial process model & Combinatorial & $0$ & $\times$ & \checkmark & \checkmark & $\times$ & $\times$ & Social & $70$ & $7000$ & $2$ & \cite{CombinatorialModel}\\  
\hline

Time-varying Markov chain clustering  & Markov & $1$ & $\times$ & \checkmark & $\times$ & $\times$ & \checkmark & Social & $6000$ & $40$ & $4$ & \cite{PlantClosure}\\ 
\hline

LCA (Latent Class Analysis) & GLM & $0$ & $\times$ & $\times$ & \checkmark & \checkmark & $\times$ & Social & $2290$ & $6$ & $32$ & \cite{LCAandSA}\\  
\hline

RLPMM (Reinforcement Learning for Policy Mixture Model) & RNN & $\infty$ & \checkmark & \checkmark & $\times$ & $\times$ & $\times$ & Clickstream & \textbf{NEI} & \textbf{NEI} & \textbf{NEI} & \cite{PolicyMixture}\\ 
\hline

HMM model choice with $BIC_H$ criteria & HMM & $1$ & $\times$ & \checkmark & $\times$ & \checkmark & \checkmark & Clickstream & $10000$ & $20$ & $10$ & \cite{ModelSelection}\\ 
\hline

Mixture of time-dependent Markov models & Markov & $\infty$ & \checkmark & \checkmark & \checkmark & $\times$ & $\times$ & Social & $3000$ & $40$ & $12$ & \cite{Nonstationarymarkov}\\ 
\hline

Markov Chain Clustering (variant) & Markov & $\infty$ & $\times$ & \checkmark & \checkmark & $\times$ & \checkmark & Social & $50000$ & $11$ & $6$ & \cite{MultinomialLogit}\\ 
\hline

Dirichlet Mixture of Hawkes Processes (DMHP) & Hawkes & $\infty$ & \checkmark & \checkmark & \checkmark & $\times$ & $\times$ & Care trajectories & $2000$ & $50$ & $5$ &  \cite{DMHP}\\  
\hline

Mixture of HMM with Gaussian or Multinomial modelling of covariates & HMM & $1$ & $\times$ & \checkmark & $\times$ & \checkmark & \checkmark & Care trajectories & \textbf{NEI} & \textbf{NEI} & \textbf{NEI} & \cite{Twostepheterogeneous}\\  
\hline

TESS & Combinatorial & $\infty$ & \checkmark & \checkmark & $\times$ & \checkmark & $\times$ & Other & $6.3 \times 10^6$ & $5$ & $6800$ & \cite{TESS}\\ 
\hline

lamb (Latent Factor Mixture Model for Bayesian Clustering) & GLM & $\infty$ & \checkmark & \checkmark & $\times$ & \checkmark & \checkmark & Clinical & $300$ & $1000$ & $10 \times 2$ & \cite{lamb}\\ 
\hline

Mixture Hidden Markov Models & HMM & $1$ & $\times$ & \checkmark & \checkmark & \checkmark & \checkmark & Social & $2000$ & $16$ & $3 \times 2 \times 2$ & \cite{seqHMM}\\  
\hline

Linear mixed-effect model & GLM & $\infty$ & \checkmark & \checkmark & \checkmark & $\times$ & \checkmark & Social & $20 000$ & $4$ & $2$ & \cite{LinearMixedEffect}\\  
\hline

Class-specific GLM & GLM & $\infty$ & \checkmark & \checkmark & \checkmark & $\times$ & \checkmark & Clinical & $750$ & $126$ & $3$ & \cite{GLMclust}\\  
\hline

Bayesian clusterwise GLM for mixed panel data & GLM & $\infty$ & \checkmark & \checkmark & \checkmark & \checkmark & \checkmark & Social & $23000$ & $4$ & $3 \times 3$ & \cite{ClusterwiseGLM}\\   
\hline
\end{longtable}%

\endgroup